\documentclass[conference,compsoc, fleqn]{IEEEtran}
\ifCLASSOPTIONcompsoc
  \usepackage[nocompress]{cite}
\else
  \usepackage{cite}
\fi
\ifCLASSINFOpdf
\else
\fi
 \usepackage[pdftex]{graphicx}

\usepackage{amsmath}
\usepackage{amssymb}
\newcommand{\FOt}{\mbox{$\mbox{\rm FO}^2$}}
\newcommand{\TGF}{\mbox{$\mbox{\rm TGF}$}}

\newcommand{\GFU}{\mbox{$\mbox{\rm GFU}$}}

\newcommand{\GFTG}{\mbox{\rm GF+TG}}
\newcommand{\GFtTG}{\mbox{\rm GF$^2$+TG}}

\newcommand{\TGFTG}{\mbox{\rm TGF+TG}}
\newcommand{\GFUTG}{\mbox{\rm GFU+TG}}
\newcommand{\GF}{\mbox{\rm GF}}
\newcommand{\GFt}{\mbox{$\mbox{\rm GF}^2$}}

\newcommand{\ExpTime}{\textsc{ExpTime}}

\newcommand{\NExpTime}{\textsc{NExpTime}}
\newcommand{\TwoExpTime}{2\textsc{ExpTime}}
\newcommand{\TwoNExpTime}{\textsc{N2ExpTime}}

\newcommand{\str}[1]{{\mathfrak{#1}}}
\newcommand{\restr}{\!\!\restriction\!\!}

\newcommand{\cI}{\mathcal{I}}

\newcommand{\type}[2]{{\rm tp}^{#1}(#2)}

\newcommand{\fh}{\mathfrak{h}}

\newcommand{\UU}{\mathsf{U}}
\newcommand{\Aux}{\mathsf{Aux}}

\usepackage{scalerel}
\newcommand{\AAA}{{\scaleobj{1.25}{\boldsymbol\alpha}}}
\newcommand{\BBB}{{\scaleobj{1.25}{\boldsymbol\beta}}}

\newtheorem{theorem}{Theorem}
\newtheorem{lemma}[theorem]{Lemma}
\newtheorem{claim}[theorem]{Claim}

\usepackage{xcolor}
\definecolor{mygreen}{rgb}{0, 0.6, 0}

\usepackage{enumitem}

\begin{document}
\title{Finite Model Theory of the Triguarded Fragment and Related Logics}

\author{\IEEEauthorblockN{Emanuel Kiero\'nski}
\IEEEauthorblockA{Institute of Computer Science\\University of Wroc\l{}aw\\
Email: emanuel.kieronski@cs.uni.wroc.pl}
\and
\IEEEauthorblockN{Sebastian Rudolph}
\IEEEauthorblockA{Computational Logic Group\\ Technische Universit\"{a}t Dresden\\
Email: sebastian.rudolph@tu-dresden.de}
}


% compsoc conferences compared to traditional conferences), use this
%Telephone: (800) 555--1212, Fax: (888) 555--1212}


\maketitle

\begin{abstract}
The Triguarded Fragment (TGF) is among the most expressive decidable fragments of first-order logic, subsuming both its two-variable and guarded fragments without equality.
We show that the TGF has the finite model property (providing a tight doubly exponential bound on the model size) and hence finite satisfiability coincides with satisfiability known to be \textsc{N2ExpTime}-complete. Using similar constructions, we also establish \textsc{2ExpTime}-completeness for finite satisfiability of the constant-free (tri)guarded fragment with transitive guards. 
\end{abstract}


% \fi
\IEEEpeerreviewmaketitle

\section{Introduction}

Ever since first-order logic (FOL) was found to have an undecidable satisfiability problem, researchers have attempted to identify expressive yet decidable fragments of FOL and pinpoint their complexity. Two of the most prominent fragments in this regard are \FOt{} (the \emph{two-variable fragment}) and \GF{} (the \emph{guarded fragment}).

For \FOt{}, decidability is retained through reducing the number of available variables to $2$, essentially restricting expressivity to independent pairwise interactions between domain elements. Decidability of \FOt{} without equality was already established in the 1960s~\cite{scott1962decision}; in the 1970s the result was extended to the case with equality~\cite{DBLP:journals/mlq/Mortimer75}. \NExpTime-completeness was established in the 1990s~\cite{GKV97}. 

\GF, which owes its decidability to the restricted ``guarded'' use of quantifiers, originated in the late 1990s~\cite{ABN98:GF}. Its satisfiability problem is \TwoExpTime-complete but drops to \ExpTime-completeness when the maximum predicate arity or the number of variables is bounded~\cite{Gra99,DBLP:journals/jolli/CateF05}. 

Both \FOt{} and \GF{} possess the \emph{finite model property} (FMP), meaning that any satisfiable sentence has a finite model. As a consequence, finite-model reasoning coincides with reasoning under arbitrary models for these fragments. 
For \FOt, existence of a finite model of only exponential size in the sentence was actually the path to establishing the above mentioned complexity. For \GF, the original FMP result gave rise to a triply exponential bound on the model size \cite{Gra99}, whereas  a tight doubly-exponential bound was established much more recently \cite{BGO14}.

In an attempt to unify \FOt{} and \GF{} toward an even more expressive decidable FOL fragment, the \emph{triguarded fragment} (\TGF) was introduced \cite{RS18}, extending prior results  \cite{Kazakov:06:Phd} as well as refining and correcting previous ideas related to ``cross products'' \cite{DBLP:conf/ijcai/BourhisMP17}. \TGF{} relaxes the guardedness restrictions of \GF{} by allowing non-guarded quantification of subformulas with up to two free variables. The price to pay for retaining decidability is that equality needs to be disallowed, or at least its use must be significantly restricted.
\TGF{} brings a new quality, as it allows one to express properties expressible in neither  \FOt{} nor \GF{}. In particular it embeds one of the most important
\emph{prefix classes}, namely G\"odel's class without equality, consisting of prenex formulas of the shape $\exists \bar{x} \forall y_1 y_2 \exists \bar{z} \varphi$.
Indeed given such a formula we can translate it to \TGF{} by eliminating the initial prefix of existential quantifiers, replacing the variables $\bar{x}$
by constants, and guarding the block of quantifiers $\exists \bar{z}$ by a dummy guard $G(y_1, y_2, \bar{z})$. Along the same lines, \TGF{} settles an open question by ten Cate and Franceschet \cite{DBLP:journals/jolli/CateF05} about the decidability of formulas of the shape $\exists \bar{x} \forall y_1 y_2 \exists \bar{z} \psi$ where $\psi$ is a guarded formula. In fact, checking satisfiability of \TGF{} is \TwoNExpTime-complete, dropping to \TwoExpTime{} when disallowing constants -- as opposed to \FOt{} and \GF, where presence or absence of constants does not make a difference, complexity-wise -- and to \NExpTime{} if the arity of predicates is bounded.

One central question left wide open in the original work on TGF \cite{RS18} is if \TGF{} has the FMP (and thus, if finite model reasoning and the associated complexity is any different from the arbitrary-model case). In that paper, it is noted that neither technique used for establishing the FMP for \FOt{} and \GF{} seems to directly lend itself for solving the question for \TGF, yet it is  conjectured that the FMP holds.  
Indeed one of this paper's core contributions is to answer this open question to the positive.

An important, practically relevant and theoretically challenging modelling feature is \emph{transitivity} of a binary relation. Neither \FOt, nor \GF, and also not \TGF{} allow for axiomatising transitivity. As a remedy, it has been suggested to provide a set of dedicated binary predicate names whose transitivity is ``hard-wired'' into the logic, that is, externally imposed by the semantics.
As it turned out, when doing so, one has to be very careful not to lose decidability. 
Unrestricted use of transitive relations in \GF{} is known to lead to undecidability 
\cite{Gra99}, this even holds for \GFt{} ($=\,$\FOt$\,\cap\,$\GF), the two-variable guarded fragment \cite{GMV99}.

One way out is to restrain the use of transitive relations so that they only are allowed to occur in guards. Indeed, satisfiability of \GFTG{} (\emph{GF with transitive guards}) was shown to be decidable and, in fact, \TwoExpTime-complete \cite{ST04}, as was -- more recently -- satisfiability of \TGFTG{} \cite{DBLP:conf/lpar/KieronskiM20}. Results for the finite model case are less extensive: so far, only finite satisfiability of \GFtTG{} was shown to be decidable and \TwoExpTime-complete \cite{KT18}. We note that \GFtTG{} does not have the FMP: indeed, a typical infinity axiom saying that, for a transitive relation $T$, every
element has a $T$-successor but is not related by $T$ to itself is naturally expressible in \GFtTG{}. We remark that all the results concerning logics with TG assume the absence of constants. It is conjectured that adding constants to the picture is technically challenging but generally possible without hazarding decidability. 

In this paper, we significantly advance the state of the art in finite model theory for the (tri)guarded fragment with and without transitive guards showing the following:

\begin{itemize}
\item %\noindent $\bullet$ 
\TGF{} (with and without constants) has the FMP, thus finite satisfiability coincides with satisfiability known to be \TwoNExpTime-complete with and \TwoExpTime-complete without constants.
\item %$\bullet$ 
Finite satisfiability of constant-free \GFTG{} (with equality) is decidable and \TwoExpTime-complete.
\item %$\bullet$ 
Finite satisfiability of constant-free \TGFTG{} (without equality) is decidable and \TwoExpTime-complete.
\item %$\bullet$ 
All three results come with a tight upper bound on the size of the finite model which is   
doubly exponential in the formula length.
\end{itemize}

The results are established through novel, rather elaborate, carefully crafted model constructions coupled with meticulous inspections of existing proofs toward the extraction of tight bounds.

\section{Logics}

We work with signatures containing relation symbols of arbitrary positive arity and, possibly, constant symbols. 
We refer to structures using Fraktur capital letters $\str{A}, \str{B}, \str{C}, \ldots$, and to their domains using
the corresponding Roman capitals $A, B, C, \ldots$. Given a structure $\str{A}$ and some $B \subseteq A$ we
denote by $\str{A} \restr B$  the restriction of $\str{A}$ to its subdomain $B$.

We usually use $a, b, \ldots$ to denote domain elements of structures, $\bar{a}$, $\bar{b},\ldots$ for tuples of domain elements, $x$, $y,\ldots$ for
variables, $\bar{x}$, $\bar{y},\ldots$ for tuples of variables, and $c$ for constants,  all of these possibly with decorations.
For a tuple of variables $\bar{x}$ we use $\psi(\bar{x})$ to denote that a formula (subformula) $\psi$ has at most free variables from $\bar{x}$.
Where convenient, tuples of elements will be treated as sets built out of its members.

For a structure $\str{A}$, a formula $\psi$ with free variables $\bar{x}$, and a tuple  $\bar{a}$ of  elements of $A$ of the same length as $\bar{x}$,
we will write $\str{A} \models \psi[\bar{a}]$ to denote that $\psi(\bar{x})$ is satisfied in $\str{A}$ under the assignment $\bar{x} \mapsto  \bar{a}$. 

For $\ell >0$, an ({\em atomic}) $\ell$-{\em type} over a finite signature $\sigma$ is a
maximal consistent set of atomic or negated atomic formulas over
$\sigma$ in $\ell$ variables $x_1, \ldots, x_\ell$ (we note that types contain equalities/inequalities and occurrences of constants if they are present in
 $\sigma$).
A \emph{type} is an $\ell$-type for some $\ell$.
We often identify a type with the formula obtained by taking the conjunction over its elements. 
 A type is \emph{guarded} if it contains a positive literal containing all its variables. Note that all $1$-types are guarded as they all contain the atom $x_1=x_1$.

Let $\str{A}$ be a structure, and let $\bar{a}$ be a tuple of its elements. 
 We denote by $\type{\str{A}}{\bar{a}}$ the unique type \emph{realized} in $\str{A}$ by the tuple  $\bar{a}$, \emph{i.e.}, the type $\alpha(\bar{x})$ such that $\str{A} \models \alpha[\bar{a}]$. We say that $\bar{a}$ is \emph{guarded} in  $\str{A}$ if $\bar{a}$ is built out of a single element or
there is a tuple of elements $\bar{b}$ containing all the elements of $\bar{a}$ and a relation symbol $P \in \sigma$
such that $\str{A} \models P[\bar{b}]$ (\emph{i.e.}, $\bar{b}$ realizes a guarded type).

We will be particularly interested in types over signatures $\sigma$ consisting of the
relation symbols (and constants, if present) used in some given formula. A particularly important role will be played by $1$-types and $2$-types.
Observe that, in the absence of constants, the number of $1$-types is bounded by a function which is exponential in $|\sigma|$, and hence also in the length
of the formula. This is because any $1$-type just corresponds to a subset of $\sigma$. On the other hand, 
when at least one constant $c$ is present, then the number of $1$-types may be doubly exponentially large.
This is because a $1$-type must completely  describe the substructure on a given element and the interpretation of $c$,
and  there are $2^{2^n}$ relations of arity $n$ on a pair of elements.

A $2$-type will be called \emph{non-degenerate} if it contains $x_1 \not= x_2$. The number of $2$-types may be doubly exponential in the length of the formula even
in the absence of constants.

Given a formula $\varphi$, its \emph{width} is the maximal number of free variables across all subformulas of $\varphi$, whereas
for a signature $\sigma$, its \emph{width} is the maximal arity among the symbols in $\sigma$.

\medskip\noindent
{\bf Guarded fragment.}
The set of \GF{} formulas is defined as the least set such that
\begin{enumerate}
	\item every atomic formula belongs to \GF{},
	\item \GF{} is closed under the standard boolean connectives $\vee, \wedge, \neg, \Rightarrow, \Leftrightarrow$, and
	\item if $\psi(\bar{x}, \bar{y}) \in$ \GF{} then $\forall \bar{x}  (\gamma(\bar{x}, \bar{y}) \Rightarrow 
\psi(\bar{x},\bar{y}))$ and $\exists \bar{x}  (\gamma(\bar{x},\bar{y}) \wedge \psi(\bar{x},\bar{y}))$
are in \GF{}, where $\gamma(\bar{x},\bar{y})$ is an atomic formula containing 
all the free variables of $\psi$.
\end{enumerate} 
The atoms $\gamma$ relativising quantifiers in point (3) of the above definition are called the \emph{guards} of the quantifiers.
For convenience, we sometimes allow ourselves to leave quantifiers for subformulas with at most one free variable to be unguarded (formally speaking, they
can be guarded by atoms $x=x$; such guards cause no problems even in those of our constructions in which equalities are generally forbidden).

In \GF{} we admit the use of equality and constants, but function symbols of arity greater than zero are forbidden.

\medskip\noindent
{\bf Triguarded fragment.}
\TGF{} is an extension of  equality-free \GF{} in which quantification for subformulas with at most two variables need not be guarded. Formally, 
the set of \TGF{} formulas is defined by taking the three syntax rules defining \GF{} formulas (substituting in them '\GF{}' to '\TGF') 
and adding the following rule:
\begin{enumerate}
	\item[4)] if
$\psi(x,y)$ is in \TGF, then $\exists x \psi(x,y)$ and $\forall x \psi(x,y)$ belong to \TGF{}.
\end{enumerate}

For convenience, instead of \TGF{} we will mostly work with the equivalent logic \GFU{}, \emph{the guarded fragment with universal role}. We assume that signatures for \GFU{}
always contain the distinguished binary relation symbol $\UU$. The set of \GFU{} formulas is then defined precisely
as the set of \GF{} formulas, but the set of admissible models is restricted to those which interpret $\UU$ 
as the universally true relation. Structures interpreting $\UU$ in this way will be called \emph{$\UU$-biquitous structures}.

It should be clear that \TGF{} and \GFU{} have the same expressive power (modulo the presence of the extra predicate $\UU$). 
For example, the \TGF-formula $\forall xy (P(x) \wedge Q(y) \Rightarrow \exists z R(x,y,z) )$ can be transformed to the (up to $\UU$) equivalent \GFU-formula
$\forall xy (\UU(x,y) \Rightarrow (P(x) \wedge Q(y) \Rightarrow \exists z R(x,y,z)) )$. In the opposite direction, \GFU{}-formulas
can be equivalently translated to \TGF{} just by appending to them the conjunct $\forall xy \UU(x,y)$, thereby axiomatising $\UU$.

In our constructions, we will frequently interpret \GFU{} formulas over non-$\UU$-biquitous structures. In this case, they are treated as usual \GF{} formulas.

\medskip\noindent
{\bf Logics with transitive guards.}  The \emph{guarded fragment with transitive guards}, \GFTG, is the logic whose formulas are constructed over purely relational
signatures containing distinguished binary symbols $T_1, T_2, \ldots$. The syntax of \GFTG{} is defined as the syntax of \GF, with the only difference
that $T_1, T_2, \ldots$ can be used only as guards. The equality symbol is allowed. Regarding the semantics, we require admissible structures to interpret $T_1, T_2, \ldots$ as transitive relations.

The \emph{triguarded fragment with transitive guards}, denoted \TGFTG, is obtained from \GFTG, as expected, by  eliminating equality, and allowing quantification for subformulas with at most two free variables to be unguarded. As in the case of \TGF, instead of \TGFTG{} we will mostly work with the equivalent logic  \GFUTG, \emph{the guarded fragment with universal role and transitive guards}, whose signatures contain the special binary symbol $\UU$.
The syntax of \GFUTG{} is as the syntax of \GFTG{}, and the set of admissible models is restricted to $\UU$-biquitous ones interpreting  $T_1, T_2, \ldots$ as transitive relations.

\medskip\noindent
{\bf Normal form.} 
We say that a \GF{} (\GFU, \GFTG, \GFUTG) formula is in \emph{normal form} if it is of the shape
\begin{align}
\nonumber \bigwedge_i \forall \bar{x}  (\gamma_i(\bar{x})  \Rightarrow \exists  \bar{y} (\gamma'_i(\bar{x}, & \bar{y})  \wedge \psi_i(\bar{x}, \bar{y}))) \\[-1ex]
\label{f:nf} \wedge& \bigwedge_j \forall \bar{x} (\gamma_j(\bar{x}) \Rightarrow  \psi_j(\bar{x}))
\end{align}
where the $\gamma_i$ and the $\gamma'_i$ and $\gamma_j$ are guards and the $\psi_i$, $\psi_j$ are quantifier-free. 
The conjuncts indexed by $i$ will be sometimes called $\forall\exists$-conjuncts, while the conjuncts indexed by $j$ will
be called $\forall$-conjuncts.
Note that in our normal form we do not explicitly include purely existential conjuncts like $\exists \bar{y} (\gamma(\bar{y}) \wedge \psi(\bar{y}))$, which 
sometimes appear in similar normal forms; nevertheless, we will occasionally allow ourselves to use them, as they can be always simulated by $\forall\exists$-conjuncts
 $\forall x (x=x \Rightarrow \exists \bar{y} (G(x,\bar{y}) \wedge \gamma(\bar{y}) \wedge \psi(\bar{y}))$, for a fresh $G$.

Let $\varphi$ be  normal form formula, $\str{A} \models \varphi$, $\zeta_i$  the $i$-th  $\forall\exists$-conjunct of $\varphi$
and  $\bar{a}$ a tuple of elements of $A$ such that $\str{A} \models \gamma_i[\bar{a}]$. We then say that a tuple $\bar{b}$ such that
$\str{A} \models \gamma'_i(\bar{a}, \bar{b}) \wedge \psi_i(\bar{a}, \bar{b})$ is a \emph{witness} for $\bar{a}$ and $\zeta_i$.

The following lemma will allow us, when dealing with (finite) satisfiability or analysing the size of minimal models of \GF{} (\GFU) or \GFTG{} (\GFUTG) formulas, to 
concentrate on 
normal form sentences of the shape as in (\ref{f:nf}). A proof of a very similar lemma can be found in \cite{ST04} (see Lemma 2 there). 

\begin{lemma}\label{l:nf}
Let $\varphi_0$ be a \GF{} (\GFU{}, \GFTG{}, \GFUTG) formula over a signature $\sigma_0$. Then one can effectively compute a set $\Delta =\{\varphi'_1, \ldots, \varphi'_d\}$
of 
normal form \GF{} (\GFU{}, \GFTG, \GFUTG) formulas over an extended signature $\sigma=\sigma_0 \cup \sigma_{\mathrm{aux}}$ of size polynomial in $|\sigma_0|$ 
 such that 
all the $\varphi'_i$ are of length polynomial in $|\varphi_0|$, $d$ is at most exponential in $|\varphi_0|$, $\bigvee_{s \le d} \varphi'_s \models \varphi_0$ and every $\str{A} \models \varphi_0$ has a $\sigma$-expansion $\str{A}' \models \bigvee_{s \le d} \varphi'_s$.
\end{lemma}

We conclude this subsection with two  simple observations with straightforward proofs allowing one to build bigger models from existing ones.
Both of them are intended to be used in the absence of constants. Their variants for the case with constants
will be presented in the Appendix. 

Let $\sigma$ be a purely relational signature. Let $(\str{A}_i)_{i \in \cI}$ be a family of $\sigma$-structures having disjoint domains.
 Their \emph{disjoint union} is the structure $\str{A}$ with
domain $A=\bigcup_{i \in \cI} A_i$ such that $\str{A} \restr A_i$ is equal to $\str{A}_i$ and for any tuple $\bar{a}$ containing elements
from at least two different $A_i$ and any relation symbol $P \in \sigma$ of arity $|\bar{a}|$ we have $\str{A} \models \neg P(\bar{a})$. 

\begin{lemma}\label{l:disunion}
Let $\varphi$ be a \GF{}  or \GFTG{} normal form formula over a purely relational signature. The disjoint union of any family of its models is also its model. 
\end{lemma}

Let $\str{A}_-$ be a $\sigma$-structure. Its \emph{doubling} is the structure $\str{A}$
built out of two copies of $\str{A}_-$. 
Formally, its domain is $A:=A_- \times \{0, 1\}$
and for each $P \in \sigma$ we set $\str{A} \models P[(a_1, \ell_1), \ldots, (a_k, \ell_k)]$ iff $\str{A}_- \models P[a_1, \ldots, a_k$] for all $a_i \in A_-$ and
  $\ell_i \in \{0, 1 \}$. 

\begin{lemma}\label{l:doubling}
Let $\varphi$ be a normal form \GF{}, \GFU{}, \GFTG{}, \GFUTG{} formula which does not use equality (or uses it only as trivial guards $x=x$) and let $\str{A}_-$ be a model of $\varphi$. Then 
its doubling $\str{A}$ is still a model of $\varphi$.
\end{lemma}

\medskip\noindent{\bf External constructions and procedures.}
In our work we will extensively use results on the complexity of guarded logics and on the size of their 
minimal finite models. We collect the relevant results in this paragraph. Generally, bounds in the original papers are
formulated in terms of the length of the input formula. We additionally give some more specific estimations,
implicit in the original works, obtained by careful yet routine analysis of the proofs. Some comments concerning these estimations can be found in the Appendix.

\begin{theorem}[~\hspace{-1.8mm}\cite{BGO14}] \label{t:gfsize}
\GF{} (with constants and equalities) has the finite model property. Every satisfiable formula has a model of size bounded doubly exponentially in its
length. More specifically, the size of minimal models of normal form formulas is bounded exponentially in the size of the signature and doubly exponentially in its width.
\end{theorem}

\begin{theorem}[~\hspace{-1.8mm}\cite{Gra99}] \label{t:gfcomp}
The satisfiability problem for \GF{} (with constants and equalities) is \TwoExpTime-complete. 
More specifically, there is a procedure that, given a normal form formula, works in time bounded polynomially in the length of the input, exponentially in the size of the signature,
and doubly exponentially in its width. 
\end{theorem}

\begin{theorem}[~\hspace{-1.8mm}\cite{KT18}] \label{t:gfttgsize}
Every finitely satisfiable \GFtTG{} formula (without constants, with equalities) has a model of size bounded doubly exponentially in its
length. More specifically, for normal form formulas, the size of their minimal finite models is  
bounded  exponentially in the number of the $\forall\exists$-conjuncts of the input and
doubly exponentially in the size of the signature.
 \end{theorem}

\begin{theorem}[~\hspace{-1.8mm}\cite{KT18}] \label{t:gfttgcomp}
The satisfiability problem for \GFtTG{} (without constants, with equalities) is \TwoExpTime-complete. 
More specifically, there is a procedure that, given a normal form formula, works in 
time bounded polynomially in the length of its input, exponentially in the number of its $\forall\exists$-conjuncts
and doubly exponentially in the size of the signature.
\end{theorem}
\section{Finite model construction for TGF (GFU)} \label{s:fmpgfu}

Let us fix a \GFU{} sentence $\varphi$ in normal form, without equality, over a purely relational signature $\sigma$ (we will explain how to cover the case
of signatures containing constants later) and let $\str{A}$ be a $\UU$-biquitous model of $\varphi$. Our goal is to build a finite $\UU$-biquitous model $\str{A}'$
of $\varphi$.

\subsection{Preparing building blocks}
Let $\AAA$   be the set of $1$-types realized in $\str{A}$. 
We construct a \GF{} $\sigma$-sentence $\varphi^*$ by appending to  $\varphi$ the following conjuncts:
\begin{align}
\label{nf2} \forall x \big(\bigvee_{\alpha \in \AAA} \alpha(x)\big)\\
\label{nf3} \bigwedge_{\alpha, \alpha' \in \AAA} \exists xy \big( \alpha(x) \wedge \alpha'(y) \wedge \UU(x,y) \wedge \UU(y,x)\big)\\
\label{nf4} \bigwedge_{P \in \sigma} \forall \bar{x} \Big( P(\bar{x}) \ \Rightarrow \!\!\!\bigwedge_{1 \le i,j \le |\bar{x}|}\!\!\! \UU(x_i, x_j)\Big)
\end{align}
saying, respectively, that only $1$-types from $\AAA$ are realized, every pair of $1$-types has a realization both-ways connected by $\UU$ 
and every guarded pair of elements is connected by $\UU$.
We can treat (\ref{nf2})--(\ref{nf4}) as normal form conjuncts.

It is clear that $\varphi^*$, treated as a \GF-formula, is satisfiable. In fact, $\str{A}$ is its model.
Thus, by the finite model  property for \GF{}, 
it also has a finite (not necessarily $\UU$-biquitous) model. We take such a finite model  $\str{C}_- \models \varphi^*$, 
and let $\str{C}$ be its doubling.
As  $\varphi^*$ does not use equality (or, to be strict, needs it only for trivial guards $x=x$, omitted from (\ref{nf2})), we have by Lemma \ref{l:doubling} that $\str{C} \models \varphi^*$. 

\begin{figure}
\begin{center}
\includegraphics[width=\columnwidth]{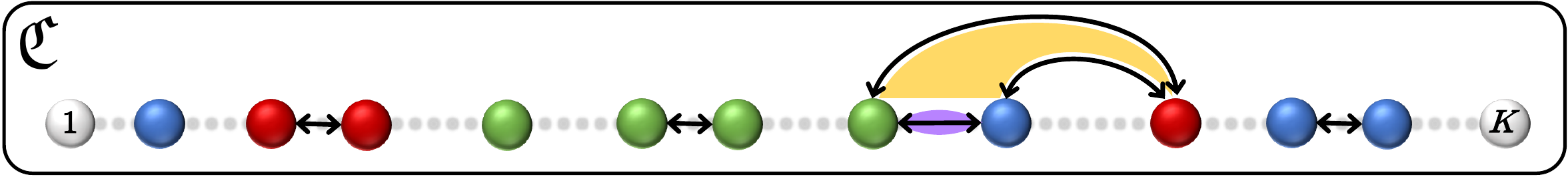}
\caption{An example structure $\str{C}$. Different colours of nodes represent different $1$-types. The black bidirectional edges depict  $\UU$-connections; the orange connection represents a ternary atom; the violet connection -- a binary one. While $\str{C}$ needs not be $\UU$-biquitous, for any pair of node colours there is a pair of distinct nodes connected by $\UU$.} \label{fig:Cstr}
\end{center}
\end{figure}

Moreover, $\str{C}$ has another convenient property.
Let us call elements $a, a' \in C$  \emph{indistinguishable} in $\str{C}$ if for any relation symbol $P \in \sigma$,
any tuple $\bar{a}_1 \subseteq C$ and any tuple $\bar{a}_2$ obtained from $\bar{a}_1$ by replacing some occurrences of $a$ by $a'$ and 
some occurrences of $a'$ by $a$ we have that $\str{C} \models P[\bar{a}_1]$ iff $\str{C} \models P[\bar{a}_2]$. Then the following holds (see Fig.~\ref{fig:Cstr}):

\begin{claim}\label{c:uconnected}
For any pair of $1$-types $\alpha, \alpha' \in \AAA$ there is a pair of their distinct realizations $a, a'$ in $\str{C}$
such that $\str{C} \models \UU[a, a'] \wedge \UU[a',a]$. Moreover, if $\alpha=\alpha'$, then we even find indistinguishable $a, a'$ with that property.    
\end{claim}
\begin{IEEEproof}
Let $b, b'$ be elements witnessing the corresponding conjunct from subsentence (\ref{nf3}) of $\varphi^*$ in $\str{C}_-$. If $\alpha\not=\alpha'$ then $b$ and $b'$ are distinct and 
we can take $a=(b, 0)$ and $a'=(b',0)$. If $\alpha=\alpha'$ then we take $a=(b,0)$ and $a'=(b, 1)$. 
By the construction of $\str{C}$, $a$ and $a'$ have the required property. Note in particular that all $1$-types in $\str{C}_-$ contain $\UU(x,x)$ as they
are realized in a $\UU$-biquitous model of $\varphi$. This implies that $\str{C} \models \UU[a,a'] \wedge \UU[a',a]$. 
\end{IEEEproof}

\medskip
From this point on, the model $\str{C}^-$ will not play any role.
However, it will be convenient to build, using Lemma \ref{l:disunion}, yet another model $\str{B} \models \varphi^*$, this time as the disjoint union of five  copies of $\str{C}$. 
Letting $K=|C|$, we assume that  the domain of $\str{B}$ is $B:=\{1, \ldots, 5K \}$; and that for $m=0,\ldots, 4$ 
 the structure on $\{mK+1, \ldots, mK+K\}$ is isomorphic to $\str{C}$.

\subsection{$\boldsymbol{\UU}$-saturation} \label{s:saturation}

We now build a finite sequence of finite structures $\str{A}_0$, $\str{A}_1, \ldots, \str{A}_f$,
each of them being a model of $\varphi^*$ and the last of them being a desired $\UU$-biquitous model $\str{A}'$
of $\varphi^*$ (and thus also of $\varphi$).

The domains of all these structures will be identical.
\begin{align}
\nonumber
A_i= B \times \{1, \ldots, 5K \} \times \{1, \ldots, 5K \}. 
\end{align}

The initial structure $\str{A}_0$ is defined 
as the disjoint union of $(5K)^2$ copies of $\str{B}$.
Namely,  for each  $k, \ell \in \{1, \ldots, 5K\}$ we make 
$\str{A}_0 \restr B \times \{k \} \times \{\ell\}$ isomorphic to $\str{B}$ 
(via the natural projection $(b,k,\ell) \mapsto b$). 
By Lemma \ref{l:disunion} we have that $\str{A}_0 \models \varphi^*$.

It is helpful to think that each of the $\str{A}_i$ is organized in a square table of size $5K \times 5K$.
In particular every cell of $\str{A}_0$ contains a copy of $\str{B}$ (which itself is a 5-fold copy of $\str{C}$), and in $\str{A}_0$, there are no connections whatsoever 
between elements from different cells.

\medskip\noindent
{\bf Outline of the construction.}  
The whole process may be seen as a careful saturation of the initial model $\str{A}_0$ with $\UU$-connections. 
In the passage from $\str{A}_i$ to $\str{A}_{i+1}$ we take a pair of distinct domain elements $b_1, b_2$
not connected by $\UU$ yet. By Claim \ref{c:uconnected}, we can find in $\str{C}$  a pair of distinct elements $a_1, a_2$ that have
the same $1$-types as $b_1, b_2$, but, in addition, are connected by $\UU$. We want to make the connection between $b_1$ and $b_2$ isomorphic to the connection between $a_1$ and $a_2$,
but after this,  $b_1, b_2$ may start to satisfy the guard $\gamma_i$ in one of the $\forall\exists$-conjuncts and thus require witnesses. To provide such witnesses we connect the pair
$b_1, b_2$ to one substructure located in one of the cells in $\str{A}_{i+1}$. Thereby, the challenge is to design a strategy which will allow us to perform a process of this kind without causing conflicts regarding the newly assigned connections.
We now propose such a strategy.

\begin{figure}
\begin{center}
\includegraphics[width=\columnwidth]{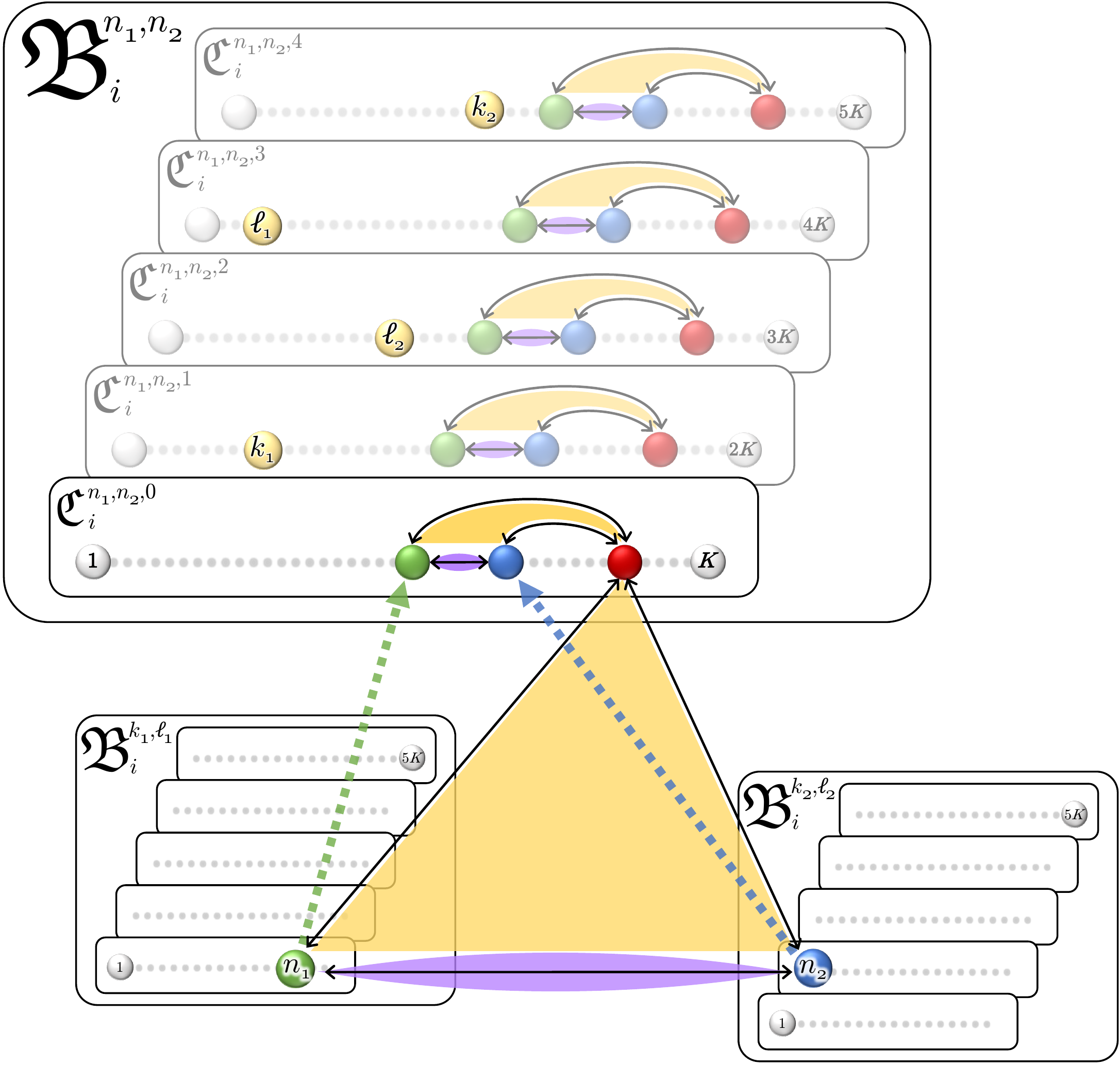}
\caption{A single step of $\UU$-saturation. The green and blue elements in $\str{B}^{n_1,n_2}_i$ are the entry elements (to the $\str{C}^{n_1,n_2,m}_0$). The orange and violet connections, as well as the black $\UU$-connections inside  $\str{C}^{n_1, n_2, 0}_i$ are already present
in $\str{C}^{n_1, n_2, 0}_0$ (isomorphic to $\str{C}$). They give rise to the orange, violet and black connections joining various cells of the table. For transparency, not all newly arising connections are shown.  \label{fig:USat}}
\end{center}
\end{figure}

\medskip\noindent
{\bf Some notation.}  To describe our strategy in detail, let us introduce some further notation. 
We denote by $\str{B}_i^{k,\ell}$ the structure in the cell $(k,\ell)$ of $\str{A}_i$, that is the structure $\str{A}_i \restr B \times \{k \} \times \{\ell \}$.
We recall that $\str{B}_0^{k,\ell}$ is isomorphic to $\str{B}$.
We will sometimes say that an element $(b,k,\ell)$ is the $b$-th element of $\str{B}_i^{k,\ell}$. 
Further, for $m=0, \ldots, 4$, we denote by $\str{C}_i^{k,\ell,m}$ the structure $\str{B}_i^{k,\ell} \restr \{mK+1, \ldots, mK+K \} \times \{k \} \times \{\ell \}$.
We recall that each $\str{C}_0^{k,\ell,m}$ is isomorphic to $\str{C}$. See Fig.~\ref{fig:USat}.

\medskip\noindent
{\bf Entry elements and their use.}
For any $1\le k,\ell \le 5K$, let $\alpha^k=\type{\str{B}}{k}$ and $\alpha^\ell=\type{\str{B}}{\ell}$. 
For each such pair $k, \ell$ we now choose a pair of \emph{entry elements} for each of the five structures in the cell $(k, \ell)$ of $\str{A}_0$, that is for the structures $\str{C}_0^{k,\ell,m}$ ($m=0, 1, \ldots, 4$).

By Claim \ref{c:uconnected}, there are distinct elements $e_1, e_2 \in C$ such that $\str{C} \models \alpha^k[e_1]
\wedge \alpha^\ell[e_2] \wedge \UU[e_1, e_2] \wedge \UU[e_2,e_1]$ and if $\alpha^k=\alpha^\ell$ then $e_1$ and $e_2$ are indistinguishable in $\str{C}$.
We choose the entry elements  $e_1^{k,\ell,m}$, $e_2^{k,\ell,m}$ to $\str{C}_0^{k,\ell,m}$ to be the corresponding copies of $e_1$ and $e_2$
in each of $\str{C}_0^{k,\ell,m}$ (recalling that the domains of all the $\str{A}_i$ are the same, the elements
$e_1^{k,\ell,m}$, $e_2^{k,\ell,m}$ belong to $C^{k, \ell, m}_i$, for all $i$). 
The entry elements will serve as a template for connecting some external pairs of elements to $\str{C}_i^{k,\ell,m}$. 
This will be done by the following construction. 

By $^+\str{C}_i^{k,\ell,m}$ we denote the structure with domain $C_i^{k,\ell,m} \cup \{b_1,b_2\}$ for some fresh elements $b_1, b_2$  such that
$^+\str{C}_i^{k,\ell,m} \restr C_i^{k,\ell,m} = \str{C}_i^{k,\ell,m}$ and 
for each $P \in \sigma$ and each tuple $\bar{a}$ containing at least one of $b_1, b_2$ we have $^+\str{C}_i^{k,\ell,m} \models P[\bar{a}]$ iff $\str{C}_0^{k,\ell,m} \models
P[\fh(\bar{a})]$, where $\fh$ is the function defined as $\fh(b_1)=e^{k,\ell,m}_1$, $\fh(b_2)=e^{k,\ell,m}_2$ and $\fh(a)=a$ for $a \in C_i^{k,\ell,m}$ (we emphasize that in this definition we copy the relations from $\str{C}_0^{k,\ell,m}$ 
and not from its possibly modified version $\str{C}_i^{k,\ell,m}$). 
In particular $^+\str{C}_i^{k,\ell,m} \models \alpha^k[b_1] \wedge \alpha^\ell[b_2] \wedge \UU[b_1, b_2] \wedge \UU[b_2, b_1]$.

\medskip\noindent
{\bf From $\boldsymbol{\str{A}_i}$ to $\boldsymbol{\str{A}_{i+1}}$.}
Assume now that the structure $\str{A}_i$ has been defined, for some $i \ge 0$.
If $\str{A}_i$ is $\UU$-biquitous then we are done. % and take $\str{A}':=\str{A}_i$.
Otherwise let $b_1, b_2$ be a pair of elements in $A_i$ such that 
$\str{A}_i \models \neg \UU[b_1, b_2]$. 
For $s=1,2$ let $k_s, \ell_s, n_s$ be such that  $b_s$  is the $n_s$-th element of  $\str{B}_i^{k_s,\ell_s}$.
Let us choose $t \in \{0, \ldots, 4\}$ such that $\str{C}_i^{n_1, n_2, t}$ does not
contain the $k_1$-th, $\ell_1$-th, $k_2$-th or $\ell_2$-th element of $\str{B}_i^{n_1, n_2}$.
Such a $t$ must exist by the pigeon hole principle. We make the structure $\str{A}_{i+1} \restr
C_i^{n_1, n_2, t} \cup \{b_1, b_2\}$ isomorphic to $^+\str{C}_i^{n_1, n_2, t}$.
The rest of the structure $\str{A}_i$ remains untouched.  Fig.~\ref{fig:USat} illustrates the described step.
Note that the orange ternary atom from $\str{C}^{n_1, n_2, 0}_i$ is inherited from $\str{C}^{n_1, n_2, 0}_0$.
Indeed a quick inspection shows that our construction never adds new local ternary atoms, where by a \emph{local atom} we mean an atom in one of the substructures $\str{C}^{k, \ell, m}_i$. Since $\str{C}^{n_1,n_2,0}_0$ satisfies (\ref{nf4}) also the black connections shown in $\str{C}^{n_1, n_2, 0}_i$ are present already in $\str{C}^{n_1, n_2, 0}_0$. 
Later we will explain that our construction never modifies guarded types, so also the violet connection is present already in $\str{C}^{n_1, n_2, 0}_0$.

\subsection{Correctness of the construction.}

We argue that for all $i$ we have $\str{A}_i \models \varphi^*$.
Note first that our construction never modifies the $1$-types of elements.
\begin{claim} \label{c:1types}
For every $i$ we have that
$\type{\str{A}_i}{a} = \type{\str{A}_{0}}{a}$.
\end{claim}
\begin{IEEEproof}
The proof goes by induction. Assume that for all $a$ and some $i$ we have that $\type{\str{A}_{i}}{a}=\type{\str{A}_{0}}{a}$.
In the passage from $\str{A}_{i}$ to $\str{A}_{i+1}$ we modify only some substructure $\str{A}_{i} \restr C_{i}^{k,\ell,m} \cup \{ b_1, b_2 \}$,
where $b_1$ is the $k$-the element of its cell and $b_2$ is the $\ell$-the element of its cell. 
By the inductive assumption they retain in $\str{A}_{i}$ their $1$-types from $\str{A}_0$ which are, $\alpha^{k}$ and $\alpha^\ell$, respectively.
In this step, we do not modify $\str{C}_{i}^{k,\ell,m}$ at all, so in particular its elements retain their $1$-types.
The $1$-type of $b_1$ ($b_2$) is set to be equal to the type of the first (second) entry element of $C_{0}^{k,\ell,m}$
which is, by our definition, of type $\alpha^k$ ($\alpha^\ell$). So also $b_1$ and $b_2$ do not change their $1$-types.
\end{IEEEproof}

\medskip
The following claim is crucial for the correctness of our construction.

\begin{claim} \label{c:noconflicts}
Let $i>0$ and assume $\str{A}_{i} \models \varphi^*$. Then every guarded tuple $\bar{a}$ of domain elements in $\str{A}_{i}$ (including the tuples guarded by $\UU$) retains its type in $\str{A}_{i+1}$, that is:
$\type{\str{A}_{i+1}}{\bar{a}}=\type{\str{A}_{i}}{\bar{a}}$. 
\end{claim}

\begin{IEEEproof}
$\str{A}_{i+1}$ is obtained from $\str{A}_i$ by making changes only in the substructure
with domain $C_i^{n_1, n_2 ,t} \cup \{b_1, b_2 \}$
(where $n_1, n_2, t, b_1, b_2$ are as in the description of the construction of $\str{A}_{i+1}$). The substructure $\str{C}_i^{n_1, n_2, t}$ itself
is not touched at all.

By the conjunct $(\ref{nf4})$ of $\varphi^*$ we have that the pair $b_1, b_2$ 
cannot be guarded in $\str{A}_i$. 
 Thus any
tuple  guarded in $\str{A}_i$ which could potentially change its type in $\str{A}_{i+1}$ must contain 
exactly one of $b_1, b_2$ and (possibly) some elements of $C_i^{n_1,n_2,t}$. Consider one such tuple $\bar{a}$.

If $\bar{a}$ is built exclusively from $b_1$ or exclusively from  $b_2$ then the claim follows
from Claim \ref{c:1types}.

Consider the case when $\bar{a}$ contains exactly one of $b_1$, $b_2$ and at least one other element.
Then $\bar{a}$ contains either elements from two different 
cells of the $5K \times 5K$ table, or elements from two different substructures $\str{C}_i^{n_1,n_2,m}$ in the cell $n_1, n_2$ (the substructure containing $b_1/b_2$ and the substructure with $m=t$); in both cases it is not guarded in $\str{A}_0$. So, its type had to be modified in the passage from
$\str{A}_j$ to $\str{A}_{j+1}$ for some $j <i-1$ (it is also possible that it was defined in several such passages;
in this case assume that $j \rightarrow j+1$ is the last of them).  

W.l.o.g.~assume that out of $b_1$, $b_2$ the tuple $\bar{a}$ contains $b_1$.
Thus the two cells which contain the elements of  $\bar{a}$ are
$(n_1, n_2)$ which contains $C_i^{n_1, n_2, t}$, and  $(k_1, l_1)$ which contains $b_1$ (as we noted, it is possible that this is actually the same cell).
Moreover,  in the structure $\str{B}_i^{n_1, n_2}$ from the cell $(n_1, n_2)$ its $k_1$-th and $l_1$-th elements are not members of $\bar{a}$, which is ensured by our choice of $t$.
Hence, by our strategy, none of the elements of $\bar{a} \setminus \{ b_1 \}$ was a member of a pair of elements which was connected
to $\str{C}_s^{k_1, l_1, m}$ for any $s$, and thus
it must be the case that the element $b_1$, together with some other element $b_3$ (having the same $1$-type as $b_2$), 
were connected to $\str{C}^{n_1, n_2, t}_j$ when forming $\str{A}_{j+1}$. If the $1$-types $\alpha^{n_1}$, $\alpha^{n_2}$ of $b_1$ and, resp., $b_3$ are different,
then  
the truth values of the atoms containing $b_1$ in $\str{A}_{j+1}$ were defined in accordance with the truth values of the tuples containing the entry element $e_1^{n_1, n_2, t}$
in the structure $\str{A}_0$, exactly as they are defined in $\str{A}_{i+1}$. So, there are no conflicts in this case. 
If $\alpha^{n_1}=\alpha^{n_2}$, it may happen that the
atoms containing $b_1$ in $\str{A}_{j+1}$ are defined in accordance with the truth values of the tuples containing the entry element $e_2^{n_1, n_2, t}$.
In this case however there are also no conflicts since the entry elements of the structures $\str{C}_0^{n_1, n_2, t}$ are 
indistinguishable 
when $\alpha^{n_1}=\alpha^{n_2}$.
\end{IEEEproof}

\medskip
By straightforward induction we get:
\begin{claim} \label{f:retaining}
Every guarded tuple of elements in $\str{A}_0$ retains its type in $\str{A}_{i+1}$.
\end{claim}

We are ready to show that $\str{A}_i \models \varphi^*$ implies $\str{A}_{i+1} \models \varphi^*$.
\begin{claim} \label{c:correctness} 
If $\str{A}_{i} \models \varphi^*$ then $\str{A}_{i+1} \models \varphi^*$. 
\end{claim}
\begin{IEEEproof}
Let us observe first that $\str{A}_{i+1} \models \varphi$. For this consider any 
$\forall \exists$-conjunct  of $\varphi$:
$\zeta=\forall \bar{x} (\gamma_i(\bar{x}) \Rightarrow \exists \bar{y} (\gamma'_i(\bar{x}, \bar{y}) \wedge \psi_i(\bar{x}, \bar{y})))$
and assume $\str{A}_{i+1} \models \gamma_i[\bar{a}]$ for some tuple $\bar{a}$. 
We consider two cases:

(a) $\str{A}_{i} \models \gamma_i[\bar{a}]$. Since $\str{A}_i \models \varphi$, we have in particular that
$\str{A}_i \models \gamma'_i[\bar{a}, \bar{b}] \wedge \psi_i[\bar{a}, \bar{b}]$ for some tuple $\bar{b}$.
As $\gamma_i'$ is an atomic formula, the tuple $\bar{a}\bar{b}$ is guarded in $\str{A}_i$ and by Claim \ref{c:noconflicts} it
retains its type in $\str{A}_{i+1}$. Hence $\str{A}_{i+1} \models \gamma'_i[\bar{a}, \bar{b}] \wedge \psi_i[\bar{a}, \bar{b}]$.
It follows that $\str{A} \models \zeta$. 

(b) $\str{A}_{i} \not\models \gamma_i[\bar{a}]$. In this case the fact $\gamma_i[\bar{a}]$
appeared first in $\str{A}_{i+1}$. Recall the construction of $\str{A}_{i+1}$ and the notation used there.
Let $\fh:\{b_1, b_2\} \cup C_i^{n_1, n_2, t} \rightarrow C_i^{n_1, n_2, t}$ be the function returning $e_s^{n_1, n_2, t}$ (for $s=1, 2$) for $b_s$ and
returning $a$ for $a \in C_i^{n_1, n_2, t}$. By the definition of $\str{A}_{i+1}$ we have that $\str{C}_0^{n_1, n_2, t} \models \gamma_i[\fh(\bar{a})]$.
As $\str{C}_0^{n_1, n_2, t} \models \zeta$ there
is a tuple $\bar{b}$ ($=\fh(\bar{b})$) in $C_0^{n_1, n_2, t}$ such that $\str{C}_0^{n_1, n_2, t} \models \gamma'_i[\fh(\bar{a}), \fh(\bar{b})] \wedge \psi_i[\fh(\bar{a}), \fh(\bar{b})]$.
Since $\gamma'_i$ is an atom,  the tuple $\fh(\bar{a}) \fh (\bar{b})$ is guarded in $\str{A}_0$. For any tuple $\bar{a}_0  \subseteq \bar{a}\bar{b}$ not containing any of  $b_1, b_2$
we have that $\bar{a}_0=$($\fh(\bar{a}_0)$) $\subseteq \fh(\bar{a})\fh(\bar{b})$, so the type of $\bar{a}_0=$ ($\fh(\bar{a}_0$) from $\str{A}_0$ is retained in $\str{A}_{i+1}$ by Claim \ref{f:retaining}.
For tuples $\bar{a}_0 \subseteq \bar{a} \bar{b}$ containing $b_1$ and/or $b_2$, their type in $\str{A}_{i+1}$ is 
the same as the type of $\fh(\bar{a}_0)$ in $\str{A}_0$, by the definition of $\str{A}_{i+1}$.
In both cases, the type of $\bar{a}_0$ in $\str{A}_{i+1}$ is the same as the type of $\fh(\bar{a}_0)$ in $\str{A}_0$. 
It follows that  $\str{A}_{i+1} \models \gamma'_i[\bar{a}, \bar{b}] \wedge \psi_i[\bar{a}, \bar{b}]$,
and thus $\str{A}_{i+1} \models \zeta$.

The reasoning for the $\forall$-conjuncts  is similar but simpler (actually, $\forall$-conjuncts are special case of $\forall\exists$-conjuncts).

As we noted the conjuncts  (\ref{nf2})--(\ref{nf4}) are normal form conjuncts and thus we can argue about them exactly as about the
conjuncts of $\varphi$.  
\end{IEEEproof}

\medskip
That our construction terminates follows from Claim \ref{c:noconflicts}. Indeed, it implies that all
pairs of elements connected by $\UU$ in $\str{A}_i$ remain connected by $\UU$ in $\str{A}_{i+1}$. On the other hand
at least one new $\UU$-connection appears in $\str{A}_{i+1}$: the one between the elements $b_1$ and $b_2$.
As the number of elements in the domain of our structures is fixed and finite, after a finite number of steps we
end up in a structure $\str{A}_f$ in which any two elements are connected by $\UU$. Since $\str{A}_0 \models \varphi^*$, Claim~\ref{c:correctness} implies,
by induction, that
$\str{A}_f \models \varphi^*$ and in particular $\str{A}_f \models \varphi$. Thus, we may take $\str{A}':=\str{A}_f$
as the desired $\UU$-biquitous model of $\varphi$.

	\subsection{Adding constants}
	The proof of the FMP for \GFU{} (\TGF{}) presented above can be extended without major problems to the case of signatures containing constants. 
	Here we outline the basic idea, for a more detailed description of the construction see Appendix \ref{a:constants}.
	
	Given a structure $\str{A}$ interpreting a signature with constants we call the subset $\hat{A} \subseteq A$ consisting of the
	interpretations of all constants the \emph{named part} of $\str{A}$. We set $\check{A} := A \setminus \hat{A}$ and call it 
	the \emph{unnamed part} of $\str{A}$. 
	
	We proceed as previously. We take a satisfiable normal form formula $\varphi$, expand it to $\varphi^*$ and take a (not necessarily $\UU$-biquitous) finite model $\str{C}_- \models \varphi^*$.
	In the absence of constants we extensively used constructions building bigger models
	out of many copies of some existing ones (Lemmas \ref{l:disunion}, \ref{l:doubling}). As this time we cannot reproduce the named part of models,
	those constructions have to be replaced by ones which multiply only their  unnamed parts.
	That is, when producing $\str{C}$ we double only $\check{\str{C}}_-$, when producing $\str{B}$ we form the disjoint union of five copies of $\check{\str{C}}$, and when producing $\str{A}_0$ we form the disjoint union of  $5K \times 5K$ copies of $\check{\str{B}}$. In each of the above steps, all the copies of the unnamed part of the input model are attached to a single, shared copy of its  named part, 
	in such a way that the restriction of the resulting structure to the union of any copy of the unnamed part and the copy of the named part is
	isomorphic to the input model. In effect, the named part of $\str{A}_0$ is inherited from the initial model
	$\str{C}_-$. It is not difficult to show that after each of the above steps we still have
	a model of $\varphi^*$. In particular $\str{A}_0 \models \varphi^*$.
	
	Next, we perform the $\UU$-saturation process. Generally, it goes as previously: we find a pair of elements $b_1, b_2$ not connected by $\UU$, join them by $\UU$ 
	and connect them to the appropriate cell of the table to provide necessary witnesses. We note only, that this time, this step involves defining the truth values of relations on tuples consisting of 
	the $b_i$, the elements from the cell to which the $b_i$ are connected and, possibly, the interpretations of constants.
		The process leads eventually to a $\UU$-biquitous model $\str{A}' \models \varphi^*$.

	\subsection{Size of models}
	We now  estimate the size of finite models that can be produced by a use of our construction. 
		
		Assume we want to  construct a finite model of a satisfiable formula $\varphi_0$ over a signature $\sigma_0$. 
		We first convert $\varphi_0$ into a disjunction of normal form formulas as guaranteed by Lemma \ref{l:nf},
		and choose a satisfiable normal form disjunct $\varphi$ (over an extended signature $\sigma$). We take an
		arbitrary model $\str{A} \models \varphi_0$. Next we 
		append to $\varphi$ the auxiliary conjuncts obtaining a normal form formula $\varphi^*$, send $\varphi^*$
		to a black box producing a finite but generally non-$\UU$-biquitous model $\str{C}_{-} \models \varphi^*$, form models $\str{C}$, $\str{B}$, $\str{A}_0$ and
		saturate $\str{A}_0$ to get finally an $\UU$-biquitous model $\str{A}'$. 
		 By Lemma \ref{l:nf}, $|\varphi|$ is polynomial in $|\varphi_0|$. $|\varphi^*|$ is
		exponential in $|\varphi|$ in the case without constants and doubly exponential in the case with constants.
		This follows from the fact that $|\varphi^*|$ contains the conjuncts (\ref{nf2}) and (\ref{nf3}) whose size
		is polynomial in the number of $1$-types over $\sigma$. So, we need to be careful and avoid
		estimating the size of $\str{A}_0$ only in terms of the length of $\varphi^*$.

   As the external black box procedure we can use any procedure constructing a finite model of a satisfiable \GF{} formula. 
	  Let as  assume that we use the model
		produced by the construction from \cite{BGO14}. 
		By Thm.~\ref{t:gfsize} the size of this model is bounded exponentially in the size and doubly exponentially in the width
		of the signature of $\varphi^*$, which is the same as the signature of $\varphi$, $\sigma$. As the size and the width of $\sigma$ are bounded by $|\varphi|$
		which, by Lemma \ref{l:nf}, is polynomial in $|\varphi_0|$, eventually  our bound on the size of $\str{C}_0$ is doubly exponential in the size of the input formula $|\varphi_0|$.
		
		Recall that $|C| = 2 |C_0|$, $|B|=5 |C| = 10 |C_0|$, and for all $i$: $|A_i|=|B|^2 \cdot |B|=(10 |C_0|)^3$ which is still doubly
		exponential in $|\varphi_0|$. Thus we get:
		
		\medskip
		\begin{theorem} \label{t:fmpgfu}
		Every satisfiable \TGF{} (\GFU{}) formula $\varphi$ (with or without constants) has a finite model of size bounded  doubly exponentially in the length of $\varphi$.
		\end{theorem}
		
		This bound is essentially optimal, since even in \GF{} wihtout constants and equality one can construct a family of satisfiable formulas $\varphi_i$, each of them of length polynomial in
		$i$, but having only models of size at least $2^{2^i}$. This is implicit in \cite{Gra99}.

		The finite model property of \TGF{} (\GFU) implies that its finite satisfiability problem is equal to its satisfiability problem and
		thus it is \TwoExpTime-complete in the absence of constants and \TwoNExpTime-complete with constants as shown in \cite{RS18}.

\section{Finite satisfiability of GF+TG and GFU+TG}

Let us recall that in case of logics with transitive guards, we work with signatures containing no constants, however, in \GFTG{}
we permit equality. Still, our decidability results for \GFUTG{} will be obtained in the absence of equality,
since, as we said, already \GFU{} with equality is undecidable.
 
For convenience, we first slightly enhance our normal form. Given a normal form \GFTG{} or \GFUTG{} formula as in (\ref{f:nf}) we  split its $\forall \exists$-conjuncts into those in which 
$\gamma'_i$ is a non-transitive symbol and those in which it is transitive. Moreover, for the latter, we assume that the guard $\gamma_i$ has only one variable.
If this is not the case -- that is we have a conjunct of the form
\begin{align}
\nonumber \forall \bar{x} \Big(\gamma_i(x_1, \ldots, x_k) \Rightarrow \exists y \big(\gamma'_i(x_j, y) \wedge \psi_i(x_j, y)\big)\Big)
\end{align}
with $k>1$ and $\gamma_i'$ using a transitive symbol -- 
we replace it by 
\begin{align}
\nonumber 
\forall \bar{x} \big(\gamma_i(x_1, \ldots, x_k) & \Rightarrow G_i^j(x_j)\big)\\  
\nonumber \wedge\ \forall x_j \Big(G_i^j(x_j) & \Rightarrow \exists y \big(\gamma'_i(x_j, y) \wedge \psi_i(x_j, y)\big)\Big)
\end{align}
where $G_i^j$ is a fresh unary symbol.

Further, we assume that all the guards $\gamma_i$  in the $\forall\exists$-conjuncts are non-transitive. If this is not the case, 
that is we have a transitive guard $\gamma_i$, say of the form $T(x,y)$, then we replace it by $G(x,y)$, for a fresh, non-transitive symbol $G$,
and append the $\forall$-conjunct $\forall xy \big(T(x,y) \Rightarrow G(x,y)\big)$. 

 Finally, for convenience, we append to normal form formulas a conjunct saying that every guarded
pair of elements is connected by $\Aux$, where $\Aux$ is a fresh binary symbol. This auxiliary conjunct does not affect satsfiability
of the formula. 

So, we will assume that normal form formulas for \GFTG{} and \GFUTG{} are of the shape:
\begin{align} 
\nonumber  \bigwedge_h \forall \bar{x} \Big(\gamma_h(\bar{x}) & \Rightarrow \exists \bar{y} \big(\vartheta_h(\bar{x}, \bar{y}) \wedge \psi_h(\bar{x}, \bar{y})\big)\Big) \\
\nonumber    \wedge \ \bigwedge_i \forall x \Big(\gamma_i(x) & \Rightarrow \exists y \big(\theta_i(x,y) \wedge \psi_i(x,y)\big)\Big)\\
\nonumber  \wedge \ \bigwedge_j  \forall \bar{x} \Big(\gamma_j(\bar{x}) & \Rightarrow \psi_j(\bar{x})\Big)\\
 \label{f:gftgnf} \wedge 
\bigwedge_{P \in \sigma} \! \forall \bar{x} \Big( P(\bar{x}) & \Rightarrow \!\!\! \bigwedge_{1 \le i,j \le |\bar{x}|} \!\!\! \Aux(x_i, x_j)\Big)
\end{align}
where the $\gamma_h$ and $\gamma_i$ are  non-transitive guards, $\gamma_j$ is a guard (transitive or non-transitive), the $\vartheta_h$ are non-transitive guards and
the $\theta_i$ are transitive guards. We recall that the transitive symbols appear in none of  $\psi_h$, $\psi_i$ and $\psi_j$. The conjuncts indexed by $h$ will be called
$\forall\exists^{ntr}$-conjuncts, the conjuncts indexed by $i$ will be called
$\forall\exists^{tr}$-conjuncts, the conjuncts indexed by $j$, together with the conjuncts speaking about $\Aux$,  will be called $\forall$-conjuncts.

\subsection{GF+TG}

Let us fix a finitely satisfiable normal form \GFTG{} formula $\varphi$ over a purely relational signature $\sigma$, of the shape as in (\ref{f:gftgnf}).
Equalities are allowed in $\varphi$.
Let $\str{A}$ be a finite model of $\varphi$. We plan to construct a finite model $\str{A}' \models \varphi$ of size
bounded  doubly exponentially in $|\varphi|$. Let $\AAA$ be the set of $1$-types realized in $\str{A}$. Let $\BBB$ be the set of non-degenerate guarded $2$-types realized in $\str{A}$. 
For $\beta \in \BBB$ let $\beta^-$ denote the set of formulas obtained from $\beta$ by removing   $T(x_1,x_2)$, $T(x_2,x_1)$, $\neg T(x_1,x_2)$, $\neg T(x_2,x_1)$,
for all transitive $T$, if they are present in $\beta$. Note that $\beta^-$ still contains the literals speaking about $T(x_1,x_1)$ and $T(x_2,x_2)$. $\beta^-$ will be called the \emph{transitive-free reduction} of $\beta$.
%Let $\BBB^-=\{ \beta^-| \beta \in \BBB \}$. 
Also, let $\str{A}^-$ denote the structure obtained from $\str{A}$ by removing all facts $T[a,b]$ for a transitive $T$ and  $a \not= b$. That is, in $\str{A}^-$ the only transitive facts may be
of the form $T[a,a]$ for some $a \in A$.

\medskip\noindent
{\bf Constructing $\boldsymbol{\str{B}^*}$ and $\boldsymbol{\str{C}^*}$.}
We now construct two auxiliary formulas out of $\varphi$.
Let 
\begin{align}
\varphi_B \ := \ \ &
\nonumber   \bigwedge_h \forall \bar{x} \Big(\gamma_h(\bar{x}) \Rightarrow \exists \bar{y} \big(\vartheta_h(\bar{x}, \bar{y}) \wedge \psi_h(\bar{x}, \bar{y})\big)\Big) \\
 \nonumber \wedge & \, \bigwedge_j \forall \bar{x} \Big(\gamma_j(\bar{x}) \Rightarrow \psi_j(\bar{x})\Big)  \\
 \nonumber \wedge &
\bigwedge_{P \in \sigma} \! \forall \bar{x} \Big( P(\bar{x}) \Rightarrow \!\!\! \bigwedge_{1 \le i,j \le |\bar{x}|} \!\!\! \Aux(x_i, x_j)\Big)\\
 \nonumber \wedge & \, \bigwedge_{T_s} \forall xy \Big(T_s(x,y) \Rightarrow x=y\Big)\\
 \nonumber \wedge & \bigwedge_{\beta \in \BBB} \exists xy \, \beta^-\hspace{-0.5pt}(x,y)\\
 \nonumber \wedge & \bigwedge_{\alpha \in \AAA} \!\! \exists x\,  \alpha(x) \,\wedge\, \forall x \!\! \bigvee_{\alpha \in \AAA} \!\! \alpha(x)
\end{align}
That is, $\varphi_B$ contains all the $\forall\exists^{ntr}$-conjuncts of  $\varphi$, all its $\forall$-conjuncts, plus the conjuncts saying that the transitive relations
do not connect distinct elements,
for every non-degenerate guarded $2$-type realized in $\str{A}$
its transitive-free reduction is realized
(note that this is a guarded formula, since $\beta^-$ remains guarded as it contains $\Aux(x,y)$), 
and that a $1$-type is realized iff it is realized in $\str{A}$. We remark that the
conjuncts speaking about $\AAA$ and $\BBB$ may contain transitive atoms outside guards, but these may only be atoms of the form $T(x,x)$ or $T(y,y)$ for some
transitive $T$.  We can replace them by $P(x)$ or, resp., $P(y)$, for some fresh $P$, and add normal form conjuncts ensuring that $\forall x (P(x) \Leftrightarrow \exists y (T(x,y) \wedge x=y))$, obtaining this way formulas
in which $T$ is used only in guard positions.  Moreover, as the restriction imposed by $\varphi_B$ on the transitive relations makes their transitivity
irrelevant, we will treat $\varphi_B$ as a \GF{} formula.

Further, let 
\begin{align} 
\nonumber
\varphi_C := 
 & \bigwedge_i \forall x \Big(\gamma_i(x) \Rightarrow \exists y \big(\theta_i(x,y) \wedge \psi_i(x,y)\big)\Big)\\
\nonumber \wedge & \bigwedge_{j: \gamma_j \text{ transitive}}  \forall \bar{x} \Big(\gamma_j(\bar{x})  \Rightarrow \psi_j(\bar{x})\Big)\\
%\nonumber \wedge \forall \bar{x} (\gamma_j(\bar{x}) \Rightarrow \psi_j(\bar{x}))\\
  \nonumber \wedge & \bigwedge_{P \in \sigma} \bigwedge_{\bar{x} \in \mathcal{S}_P} \forall xy \Big( P(\bar{x}) \Rightarrow \Aux(x,y)\Big)\\
  \nonumber \wedge & \ \forall xy \Big(\Aux(x,y) \Rightarrow (x \not=y \Rightarrow \bigvee_{\beta \in \BBB} \beta^-(x,y))\Big)\\
  \nonumber \wedge & \bigwedge_{\alpha \in \AAA} \exists x \alpha(x) \wedge \forall x \bigvee_{\alpha \in \AAA} \alpha(x)
\end{align}
where $\mathcal{S}_P$ is the set of tuples of length equal to the arity of $P$ built out of variables $x$ and $y$, and containing at least one
occurrence of each of them.

That is, $\varphi_C$ contains all the $\forall\exists^{tr}$-conjuncts and those $\forall$-conjuncts of $\varphi$ that do not speak about $\Aux$, plus the conjuncts saying that 
for every guarded tuple built out of two elements these two elements are connected by $\Aux$, every guarded pair
of distinct elements satisfies the transitive-free reduction of a guarded $2$-type from $\str{A}$, 
and that a $1$-type is realized iff it is realized in $\str{A}$.
 Note that the $\forall$-conjuncts we include here have transitive $\gamma_j$, so they
use at most two variables; we may assume that they are $x, y$.
The remaining conjuncts also use only variables $x$ and $y$, so $\varphi_C$ is a formula belonging to \GFtTG{} (again after the appropriate adjustments concerning the use of transitive relations
in each of the $\alpha$ and $\beta$).

Note that both $\varphi_B$ and $\varphi_C$ are finitely satisfiable, as the former is satisfied in $\str{A}^-$ and the latter in $\str{A}$. 
Treating $\varphi_B$ as a \GF{} formula we take its small finite model $\str{B}$ 
as guaranteed by \cite{BGO14}. Similarly, treating $\varphi_C$ as a \GFtTG{} formula
we take its small finite model $\str{C}$, 
as guaranteed by \cite{KT18}. 
We remark here that while \cite{KT18} considers explicitly only signatures with relation
symbols of arity $1$ and $2$, a routine inspection shows  that the constructions there work smoothly even if symbols of arity greater than $2$ are allowed,
which is the case in our work.
Indeed, the presence of relations of arity greater than $2$ could be important in \cite{KT18} only when  $2$-types are assigned to pairs of elements. 
However, those $2$-types are read off from a pattern model, and  whether they contain higher arity relations or not is not relevant.
 Since $\varphi_C$ is a two-variable formula, we may assume that $\str{C}$ contains no fact
with more than two distinct elements.  We note that  $\str{C}$ happens to satisfy all the $\forall$-conjuncts of $\varphi$. 
The conjuncts with transitive $\gamma_j$ are included explicitly in $\varphi_C$; for those with non-transitive $\gamma_j$ assume that 
$\str{C} \models \gamma_j[\bar{a}]$ for some tuple $\bar{a}$. By our assumption on $\str{C}$ the tuple $\bar{a}$ is built out
of at most two elements. So, either it uses only one element and then its $1$-type is realized in $\str{A}$ (by the conjunct of
$\varphi_C$ speaking about $\AAA$), or it uses two elements, 
and then the  transitive-free reduction of their $2$-type is the same as the reduction of some $2$-type realized in $\str{A}$ (by the
conjunct speaking about $\BBB$). 
Since $\str{A} \models \varphi$, and in particular $\str{A} \models \forall \bar{x} (\gamma_j(\bar{x}) \Rightarrow \psi_j(\bar{x}))$,
and recalling that $\psi_j$ does not contain any transitive relations,  it follows that $\str{C} \models
\psi_j[\bar{a}]$. 

By the conjuncts of  $\varphi_B$ and $\varphi_C$ speaking about $\AAA$, we know that the sets of $1$-types realized in $\str{B}$ and $\str{C}$ are equal (concretely, they are equal to
$\AAA$). 
We now construct models  $\str{B}^* \models \varphi_B$ and $\str{C}^* \models \varphi_C$ such that the number of realizations of $\alpha$ in $\str{B}^*$ 
 is equal
to the number of its realizations in $\str{C}^*$, for all  $\alpha$. 

Let $m$ be the maximal number of realizations of a $1$-type $\alpha$ in $\str{B}$ (over all $\alpha \in \AAA$). 
Let $\str{C}^*$ be the disjoint union of $m$ copies of $\str{C}$. By Lemma \ref{l:disunion} we have that $\str{C}^* \models \varphi_C$.
Note that for every $\alpha$ the number of realizations of $\alpha$ in $\str{B}$ is less than or equal to
the number of its realizations in $\str{C}^*$. To make these numbers equal we successively adjoin additional realizations
of the appropriate $1$-types to $\str{B}$.
This is simple: to adjoin a realization $b$ of a $1$-type $\alpha$ we choose a pattern element $a$ of type $\alpha$ in $\str{B}$
and make the structure on $\big(B \setminus \{a \}\big) \cup \{b\}$ isomorphic to $\str{B}$; we add no facts containing both $a$ and $b$
to the structure. After each such step the resulting structure is still a model of $\varphi_B$.
This way we eventually get the desired $\str{B}^*$. 

\medskip\noindent
{\bf Constructing $\boldsymbol{\str{D}}$.} Let $K=|B^*|=|C^*|$. Let $b_0, \ldots, b_{K-1}$ be any enumeration of the elements of $B^*$ and
let $\alpha_i =\type{\str{B^*}}{b_i}$, for all $0 \le i < K$. We create a new structure $\str{D}$, with domain $D=\{0, \ldots, K-1\} \times \{0 \ldots, K-1\}$.
We set $\type{\str{D}}{k,\ell}:=\alpha_{k+\ell \mod K}$. Viewing in the natural way $\str{D}$ as a square table, we see that from its every row and its every column
one can construct bijections into $\str{B}^*$ and $\str{C}^*$ preserving the $1$-types. Without modifying the $1$-types, we can thus  define the  structure of  $\str{D}$ on every row and every
column in such 
a way that they become  isomorphic copies of $\str{B}^*$, and $\str{C}^*$, respectively. This completes the definition of $\str{D}$, that is, we add no further facts to it. 
Note that transitive relations cannot connect elements from different columns.

Call a guarded tuple of elements of $\str{D}$ \emph{vertical} (\emph{horizontal}) if it belongs to a single column (row) of the table.
Observe that the tuples built out of a single element are both vertical and horizontal and every guarded tuple in $\str{D}$ is either vertical or horizontal
by the definition of $\str{D}$.  

Note that $\str{D}$ satisfies the $\forall\exists^{tr}$-conjuncts of $\varphi$.
Indeed, any element $a$ satisfying some $\gamma_i[a]$ has the required witness in its column, since the relevant
conjunct is a member of $\varphi_C$. Also the $\forall$-conjuncts are satisfied: we have explained that they are satisfied in $\str{C}$, so it follows that they are
also satisfied in $\str{C}^*$, and thus in every column; on the other hand $\varphi_B$ includes them explicitly, so they are safisfied in every row. 
  Concerning
the $\forall\exists^{ntr}$-conjuncts,  note that any horizontal guarded tuple $\bar{a}$ satisfying any of $\gamma_h[\bar{a}]$ has the required witnesses
in its row (since its row satisfies $\varphi_B$). 

The only problem is that some vertical guarded tuples may not have witnesses for some
$\forall\exists^{ntr}$-conjuncts. Note that every such tuple is
built out of precisely two elements. Indeed, by our assumption about $\str{C}^*$ there are no vertical guarded tuples containing three or more distinct elements there, and on the other hand any tuple  built out of a single element  has its witnesses in its row.

We will fix the above problem by taking an appropriate number of copies of $\str{D}$ and adjoining every vertical guarded pair of elements  to a row in some different copy of $\str{D}$.  This will be done in a circular way, reminiscent of the
small model construction for \FOt{} from \cite{GKV97}.  
We emphasise that this process is much simpler than the $\UU$-saturation process from Section \ref{s:saturation} since this time
we do not need to deal with pairs of elements from different copies of our basic building block $\str{D}$. Let us turn to details.

\medskip\noindent
{\bf Building a small model of $\boldsymbol{\varphi}$.} Assume that $\str{D}_1$ and $\str{D}_2$ are two copies of $\str{D}$. Consider a 
vertical guarded pair of distinct elements $b_1, b_2$ in  $\str{D}_1$. As the column of this pair is a model of $\varphi_C$ we know that 
it satisfies $\Aux[b_1, b_2]$ and thus also it satisfies $\beta^-[b_1, b_2]$ for some $\beta \in \BBB$. 
(We recall that $\beta^-$ does not mention any transitive connection between $b_1$, $b_2$ but such connections may
be present in $\str{D}_1$.) 
As any row  $\str{E}$ of $\str{D}_2$ is a model of $\varphi_B$ it follows that
this row contains a pair of distinct elements $a_1, a_2$ such that $\str{E} \models \beta^-(a_1, a_2)$. 
(Here, there are no transitive connections between $a_1$ and $a_2$ as ensured by $\varphi_B$.)

We now describe the procedure to which we will later refer by saying:  \emph{we connect the pair $b_1, b_2$ to the row $\str{E}$ (using $a_1, a_2$
as a template)}.
 For any tuple $\bar{a}$ containing at least one of $b_1, b_2$ and
some  elements of $E \setminus \{a_1, a_2\}$ and any non-transitive relation symbol $P \in \sigma$ of arity $|\bar{a}|$ we add the fact $P[\bar{a}]$ iff 
$\str{E} \models P[\fh(\bar{a})]$ where $\fh$ is the function returning $a_s$ for $b_s$ ($s=1,2$) and $a$ for all $a \in E$. In other words, we connect $b_1, b_2$ with the elements of $E \setminus \{a_1, a_2\}$
exactly as $a_1$, $a_2$ are connected with these elements in $\str{E}$. We add no other facts. (In particular in this procedure we add no transitive connections.) This way the guarded tuples containing any of $b_1$, $b_2$ (or both) and possibly some elements of $E$, have all the required witnesses for the $\forall\exists^{ntr}$-conjuncts since the structure we have defined on $\big(E \setminus \{a_1, a_2\}\big) \cup \{b_1, b_2\}$ is isomorphic to $\str{E}$ when the transitive relations are not taken into account (and recall that the $\forall\exists^{ntr}$-conjuncts do not mention transitive relations at all).

Now, let the structure $\str{A}'$ be the disjoint union of $3K$ copies of $\str{D}$. 
More specifically its domain is $A'=\{0,1,2\} \times \{0, \ldots, K-1 \} \times D$
and $\str{A}' \restr \{i \} \times \{j\} \times D$ is isomorphic to $\str{D}$ for all $i,j$.
Denote $A_i=\{i \} \times \{0, \ldots, K-1\} \times D$, for $i=0,1,2$.

For every vertical guarded pair of distinct elements $b_1, b_2$ in $A_i$ chose a row in a copy of $\str{D}$ contained in $A_{(i+1) \hspace*{-3pt}\mod 3}$ such that this row has not yet been used by any other pair
from the column of $b_1$ and $b_2$ and connect $b_1$, $b_2$ to this row (using an appropriate template).  As the number of pairs of elements
in the row of $b_1, b_2$ is smaller than $K^2$ and there are $K$ copies of $\str{D}$ in each of the $A_i$, and each of them has $K$ rows,
we have sufficiently many rows to perform this step.

Using three sets $A_i$ and applying the above described circular strategy guarantee that the process can be performed without
conflicts: if an element $b$ is connected to a row $\str{E}$ then no element of $\str{E}$ is ever connected to the row of $b$. 

Now it should be clear that indeed the eventually obtained structure models $\varphi$.

\medskip\noindent
{\bf Size of models and complexity.} 
We now analyse the small model construction described in the previous paragraphs and obtain the following (optimal) bound
on the size of minimal finite models for \GFTG{}. For further purposes in the second part of this theorem we formulate
a more specific bound for normal form formulas.  

\begin{theorem} \label{t:smallmodelGFTG}
Every finitely satisfiable \GFTG{}  formula without constants has a model of size bounded doubly exponentially in its length.
For finitely satisfiable normal form formulas there are models of size bounded  exponentially in the size of the signature and the number of their $\forall\exists$-conjuncts, and doubly exponentially in the width of the signature.
\end{theorem} 
\begin{IEEEproof}
Let us  summarize the steps needed to produce a small model of an input finitely satisfiable \GFTG{} sentence $\varphi_0$ over a signature $\sigma_0$. We convert it into a disjunction of normal form formulas  over an extended
signature $\sigma$ as in
Lemma \ref{l:nf}, and choose its finitely satisfiable normal form disjunct $\varphi$ and its finite model $\str{A}$.
Let $r$ be the number of symbols in $\sigma$ and $w$ its width. As by Lemma \ref{l:nf} $|\varphi|$ is polynimial in $|\varphi_0|$, both $r$ and $w$ are
polynomial in $|\varphi_0|$. 
Perform our small model construction for $\varphi$ and let $\varphi_B$,
$\varphi_C$,  $\str{B}$, $\str{C}$, $\str{B}^*$, $\str{C}^*$, $\str{D}$ and $\str{A}'$ be as in this construction.

Recall that as $\str{B}$ we take the small model for $\varphi_B$ (which is a normal form formula over the signature $\sigma$), constructed as in \cite{BGO14}. By Thm.~\ref{t:gfsize} the size of $\str{B}$ is bounded exponentially in $r$ and doubly exponentially in $w$, that is doubly exponentially in $|\varphi_0|$.

Concerning $\str{C}$, we take as it the small model for $\varphi_C$ (which is a normal form formula over the signature $\sigma$), constructed 
by applying the small model construction from  \cite{KT18}. By Thm.~\ref{t:gfttgsize}, $|C|$ is bounded 
exponentially in
the number of its $\forall\exists$-conjuncts (which are actually taken from $\varphi$) and doubly exponentially
the size of the signature. So, it is bounded doubly exponentially in $|\varphi_0|$. 

Further, each of the structures $\str{B}^*$, $\str{C}^*$ has size at most $|B| \cdot |C|$, the structure $\str{D}$ -- at most
$|B^*|^2$ and the final model $\str{A}'$ -- at most $3|B^*||D|=3|B^*|^3=3|B|^3|C|^3$, which is  still doubly exponential  in  $|\varphi_0|$. 

The second part of the theorem, concerning normal form formulas follows easily from the information on the size of $\str{C}$ and $\str{B}$
given above and from the observation that the final estimation on $|A'|$ is polynomial in $|B|$ and $|C|$.
\end{IEEEproof}

\medskip
Thm.~\ref{t:smallmodelGFTG} immediately yields a \TwoNExpTime-upper bound on the complexity of the finite satisfiability problem for \GFTG{}: it suffices
to guess a bounded size structure and verify that it is indeed a model of the input formula. To get the optimal \TwoExpTime-upper complexity bound, instead of guessing a model, we may construct $\varphi_B$ and $\varphi_C$ for various sets $\AAA$ and $\BBB$ and
test their satisfiability. We first make two observations, Lemma \ref{l:alg} and Lemma \ref{l:gftg2types}, the second of which reduces the number of possible choices
of $\BBB$, which will be crucial for lowering the complexity.

\begin{lemma} \label{l:alg}
Let $\varphi$ be a normal form \GFTG{} formula over a signature $\sigma$. Then 
$\varphi$ is finitely satisfiable iff there are sets 
$\AAA$, $\BBB$ of $1$-types, and, resp., non-degenerate guarded $2$-types over $\sigma$, such that (i)  the formulas $\varphi_B$ and $\varphi_C$ constructed 
with such $\AAA$ and $\BBB$ have
finite models, (ii)  for every $\beta \in \BBB$ any two-element structure of type $\beta$ satisfies all the $\forall$-conjuncts of $\varphi$.
\end{lemma}
\begin{IEEEproof}
$\Rightarrow$ Assume $\varphi$ has a finite model $\str{A}$. As $\AAA$ and $\BBB$ take the set of $1$-types and, resp., non-degenerate guarded $2$-types realized in $\str{A}$. 
Then (i) holds since $\varphi_B$ is satisfied in $\str{A}^-$ and $\varphi_C$--- in $\str{A}$ (cf.~the paragraph about the construction of $\str{B}^*$ and
$\str{C}^*$), and (ii) holds since the guarded types in $\BBB$ are taken from a model of $\varphi$. 
$\Leftarrow$ Having $\AAA$ and $\BBB$ 
satisfying (i) and (ii) we can perform a finite model construction for $\varphi$ exactly as we did in this section (with condition (ii) used to 
ensure that $\str{C}$ respects all the $\forall$-conjuncts of $\varphi$). 
\end{IEEEproof}

\medskip

\begin{lemma} \label{l:gftg2types}
If $\varphi_C$  has a finite model then it has one in which the number of realized
$2$-types is bounded polynomially in the number of $1$-types, the number of the $\forall\exists$-conjuncts of $\varphi$ 
and the number of transitive relations.
\end{lemma}
\begin{IEEEproof}
Let $\str{C}$ be a finite model of $\varphi_C$. Recall that $\varphi_C$ is in \GFtTG{} and thus, as previously,  we may assume that $\str{C}$ contains no facts with more than two distinct elements. Additionally, we may also assume that every pair of distinct elements in $\str{C}$ is
connected by at most one transitive relation (this condition is ensured in the finite model construction in \cite{KT18}; models satisfying this
condition are called \emph{ramified} there).
Let $\BBB$ be the set of $2$-types realized in $\str{C}$. 
For a $2$-type $\beta$, by $\beta \restr x_1$ ($\beta \restr x_2$)
we denote the subset of $\beta$ consisting of those literals which use only variable $x_1$ ($x_2$); similarly by $\beta \restr \sigma_{tr}$ we
denote the subset of $\beta$ consisting of those literals which use a transitive symbol. 
Let $\beta^{-1}$ be the result of switching the variables in $\beta$.

Let us introduce an equivalence relation $\sim$ on $\BBB$ as follows: $\beta_1 \sim \beta_2$
iff the following conditions hold (i) $\beta_1 \restr x_1 = \beta_2 \restr x_1$, $\beta_1 \restr x_2 = \beta_2 \restr x_2$, $\beta_1 \restr \sigma_{tr}  = \beta_2 \restr 
\sigma_{tr}$,
(ii)  for each conjunct $\forall x (\gamma_i(x) \Rightarrow \exists y (\theta_i(x,y) \wedge \psi_i(x,y)))$ it holds that 
 $\beta_1 \models \theta_i(x,y) \wedge \psi_i(x,y)$ iff $\beta_2 \models \theta_i(x,y) \wedge \psi_i(x,y)$
and $\beta^{-1}_1 \models \theta_i(x,y) \wedge \psi_i(x,y)$ iff $\beta^{-1}_2 \models \theta_i(x,y) \wedge \psi_i(x,y)$. Observe that $\beta_1 \sim \beta_2$ iff
$\beta^{-1}_1 \sim \beta^{-1}_2$.

In every equivalence class of $\sim$ we distinguish  one of its members.  We do this in such a way that if $\beta$ is distinguished in its class then $\beta^{-1}$ 
is also distinguished in its class. For every pair of elements $a, b \in C$, if its type $\type{\str{C}}{a, b}$ is not distinguished
in its class, change this type to the one which is distinguished there. The strategy of distinguishing always both $\beta$ and $\beta^{-1}$
allows us to perform this process without conflicts which could potentially arise when the types of  pairs $a,b$ and $b,a$ are defined.

In so obtained structure $\str{C}'$ the interpretation of the transitive symbols remains unchanged (so they all remain transitive)
and every element has precisely the same witnesses for every $\forall \exists$-conjunct as it has in $\str{C}$ (even thought it may
be connected to them by different $2$-types). The $\forall$-conjuncts are satisfied in 
$\str{C}'$ as all the types are imported from $\str{C}$ which is a model of $\varphi_C$. Thus, still  $\str{C}' \models \varphi_C$.

It is readily verified that the number of equivalence classes of $\sim$ is bounded polynomially in the number of $1$-types, in the number of the $\forall\exists$-conjuncts,
and in the number of transitive relations (the latter follows from the fact that $\str{C}$ is ramified). From this
we get that $\str{C}'$ is as required.
\end{IEEEproof}

\begin{theorem} \label{t:gftgcomplexity}
The finite satisfiability problem for \GFTG{} without constants is \TwoExpTime-complete.  
For normal form formulas it works in time polynomial in the size of the input formula, exponential in the number of the $\forall\exists$-conjuncts, and doubly exponential
in the size and width of the signature.
\end{theorem}
\begin{IEEEproof}
The lower bound is inherited from $\GF$ \cite{Gra99} or from \GFtTG \cite{KT18}. Let us  justify the upper bound.
Let $\varphi_0$ be any \GFTG{} formula. Convert it into a disjunction of normal form formulas over a signature $\sigma$ as in Lemma~\ref{l:nf} and test satisfiability of each  of its disjuncts. 
To do the latter, for a single normal form disjunct $\varphi$ over a signature $\sigma$ construct all possible sets of $1$-types $\AAA$, and all
possible sets of guarded non-degenerate $2$-types of cardinality 
bounded as in Lemma~\ref{l:gftg2types}. 
For each pair of such $\AAA$ and $\BBB$  construct $\varphi_B$ and $\varphi_C$, which are normal form formula, and test their finite satisfiability using the algorithm from Thm.~\ref{t:gfcomp}, and, resp., Thm.~\ref{t:gfttgcomp}. Our algorithm returns 'yes' iff for some normal form disjunct $\varphi$ and some choice of $\AAA$ and $\BBB$ 
both the external algorithms return 'yes'.

The correctness of the algorithm follows from Lemmas \ref{l:nf}, \ref{l:alg} and \ref{l:gftg2types}.

Denote
$h$ the number of the $\forall\exists$-conjuncts of $\varphi$,
$k$ the number of transitive relations,
$r$ the size of $\sigma$ and $w$ its width.

Recall that the number of disjuncts in the normal form of $\varphi_0$ is at most exponential in $|\varphi_0|$.
Note that the number of $1$-types is $2^r$, so the number of possible choices for $\AAA$ is $2^{2^r}$.
Due to Lemma \ref{l:gftg2types} we can restrict attention to sets $\BBB$ of size bounded
polynomially in $2^r$, $h$ and $k$. Observing that the number of $2$-types is $2^{\mathcal{O}(r \cdot 2^w)}$ 
we see that the number of relevant choices of $\BBB$ is doubly exponential in $r$ and $w$
and singly exponential in $h$ and $k$.

By Thm.~\ref{t:gfcomp} the first of the external procedures works in time 
polynomial in $|\varphi_B|$, exponential in $r$ and doubly exponential in $w$.
By Thm.~\ref{t:gfttgcomp} the second procedure works in time polynomial
in $|\varphi_C|$, exponential in $h$ and doubly exponential in $r$.

Regarding the size of $\varphi_B$ and $\varphi_C$ they are both bounded
polynomially in $|\varphi|$, and in the size of $\AAA$ and $\BBB$,
that is they are doubly exponential in $r$ and $w$ and singly exponential in $h$ an $k$.

Gathering the above, the claim for normal form formulas follows. The upper bound for arbitrary formulas
follows from the fact that $|\varphi|$ is polynomial in $|\varphi_0|$, and thus
all the parameters $r$, $w$, $h$ are polynomial in $|\varphi_0|$.
\end{IEEEproof}

\subsection{GFU+TG} 

Assume that $\varphi$ is a finitely satisfiable \GFUTG{} normal form formula without equality and let $\str{A}$ be its finite ($\UU$-biquitous) model. We will explain how to construct a
($\UU$-biquitous) model $\str{A}'$ of $\varphi$ of size bounded doubly exponentially in $|\varphi|$. The whole construction
is almost identical to the construction of a small model of a satisfiable \GFU{} normal form formula in Section \ref{s:fmpgfu}. There are only two, rather natural,
differences: first, obviously, we use a different black box procedure; second, when constructing  the $^+\str{C}_i^{k,\ell,m}$ structures we must properly handle the transitive
relations.

So, we first construct the auxiliary formula $\varphi^*$ exactly as in Section \ref{s:fmpgfu} (we only need the adjustment concerning transitive atoms of
the form $T(x,x)$ used outside the guards in the $\alpha(x)$, similar to that for $\varphi_B$ and $\varphi_C$), and  treating it as a $\GFTG$ formula we take its small (not necessarily $\UU$-biquitous) model $\str{C}_-$ as guaranteed by Thm.~\ref{t:smallmodelGFTG}.
We proceed as in Section \ref{s:fmpgfu}, building the doubling $\str{C}$ of $\str{C}_-$, a $5$-fold copy $\str{B}$ of $\str{C}$, and 
a $5K \times 5K$ table $\str{A}_0$ of copies of $\str{C}$, where  $K=|C|$. All these structures are models of $\varphi^*$ by Lemmas \ref{l:doubling} and \ref{l:disunion}.

We employ the same notation as in the case of \GFU{}. We choose the entry elements for the structures $\str{C}_0^{k, \ell, m}$ and proceed
to the definition of the structures $^+\str{C}_i^{k,\ell,m}$. As previously, $\str{C}_i^{k, \ell, m}$  is the structure with  domain $C_i^{k,\ell,m} \cup \{b_1,b_2\}$ for some fresh elements $b_1, b_2$  such that $^+\str{C}_i^{k,\ell,m} \restr C_i^{k,\ell,m} = \str{C}_i^{k,\ell,m}$. Concerning the connections involving the new elements
$b_1, b_2$, for each {\bf non-transitive} $P \in \sigma$ and each tuple $\bar{a}$ containing at least one of $b_1, b_2$ we set that $^+\str{C}_i^{k,\ell,m} \models P[\bar{a}]$ iff $\str{C}_0^{k,\ell,m} \models
P[\fh(\bar{a})]$, where $\fh$ is the function defined as $\fh(b_1)=e^{k,\ell,m}_1$, $\fh(b_2)=e^{k,\ell,m}_2$ and $\fh(a)=a$ for $a \in C_i^{k,\ell,m}$. That is, as previously,  we copy the relations from $\str{C}_0^{k,\ell,m}$, but only those non-transitive. In particular the elements $b_1, b_2$ remain not connected by any transitive relation even if the entry elements are connected by some in $\str{C}_0^{k,\ell,m}$. 

We then build successively the structures $\str{A}_1$, $\str{A}_2$, $\ldots$ exactly as in the case of \GFU, obtaining finally a $\UU$-biquitous structure
$\str{A}_f$ which we take as the desired $\str{A}'$.

The correctness of the construction can be proved as in the case of \GFU. The analogues of the Claims \ref{c:1types}, \ref{c:noconflicts} and \ref{f:retaining}
can be proved with literally no changes. Also, the proof of Claim \ref{c:correctness} is almost the same.  As we have emphasized, in our process we do not add any transitive
connections; we also do not modify any $2$-types containing any transitive connections. So, the $\forall$-conjuncts with transitive guards are satisfied in all the $\str{A}_i$. Note also that 
we never need new witnesses for the $\forall\exists^{tr}$-conjuncts, as such witnesses are required only for tuples built out of a single element,
and such tuples have the required witnesses already in $\str{A}_0$. Thus, we only need to take care of witnesses for the conjuncts not mentioning 
transitive relations, and for satisfaction for the $\forall$-conjuncts with non-transitive guards. For such conjuncts the fact that we do not copy from $\str{A}_0$ the complete types of tuples, but rather their 
transitive-free parts is not relevant.

\medskip\noindent
{\bf Size of models and complexity.}
The following theorem follows from an analysis of the size of the $\str{A}_i$ structures, similar to that in the case of \GFU{}, and a use of the estimation on
the size of $\str{C}_-$ from the second part of Thm.~\ref{t:smallmodelGFTG}.
\begin{theorem}
Every finitely satisfiable \GFUTG{} (\TGFTG) formula without constants has a model of size bounded doubly exponentially in its length.
\end{theorem} 

Concerning the complexity, we first make the following observation.

\begin{lemma} \label{l:tgftgtest}
A normal form \TGFTG{} formula $\varphi$ without constants is finitely satisfiable iff there exists a set of $1$-types $\AAA$ such that
the formula $\varphi^*$, treated as a \GFTG{} formula, has a finite (not necessarily $\UU$-biquitous) model. 
\end{lemma}
\begin{IEEEproof}
$\Rightarrow$ If $\str{A}$ is a finite model of $\varphi$ then we take as $\AAA$ the set of $1$-types realized in $\str{A}$,
and note that $\str{A} \models \varphi^*$.  $\Leftarrow$ Given a finite (not necessarily $\UU$-biquitous) model $\str{C}_- \models \varphi^*$ we construct
a $\UU$-biquitous model of $\varphi^*$ (and thus also of $\varphi$) as described above.
\end{IEEEproof}

\medskip
Finally we get:
\begin{theorem}
The finite satisfiability problem for \GFUTG{} (\TGFTG) without  constants is \TwoExpTime-complete.  
\end{theorem}
\begin{IEEEproof}
The lower bound is inherited from \GFTG. To justify the upper bound we design the following algorithm:
For input $\varphi_0$ we convert it into normal form over a signature $\sigma$ and for each of its disjuncts  $\varphi$
and each possible choice of $\AAA$ construct  $\varphi^*$ and test its finite satisfiability  using 
the procedure for \GFTG. We answer 'yes' iff at least one of the tests is positive.

The correctness of this algorithm follows from Lemmas \ref{l:nf} and  \ref{l:tgftgtest}. Recall that the number of 
disjuncts in normal form of $\varphi_0$ is exponential in $|\varphi_0|$ and the number of choices of $\AAA$ is doubly exponential
in $|\sigma|$ and thus also in $|\varphi_0|$. By Thm.~\ref{t:gftgcomplexity} a single finsat test for $\varphi^*$
takes time polynomial in $|\varphi^*|$ (exponential in $|\varphi_0|$), exponential in the number of the $\forall\exists$-conjuncts and doubly exponential in the size and width of $\sigma$ (which are polynomial in $|\varphi_0|$). So, overall, the algorithm  is doubly exponential in $|\varphi_0|$.  
\end{IEEEproof}

\section{Conclusion}

Settling an open problem, we established the finite model property of the triguarded fragment 
(and consequently of guarded formulae preceded by a sequence $\exists^* \forall \forall \exists^*$ of unguarded quantifiers \cite{DBLP:journals/jolli/CateF05}),
even providing a doubly exponential upper bound on the model size. Using similar ideas, we settled open problems concerning the guarded and triguarded fragment extended by transitive guards, providing tight complexity bounds for their finite satisfiability problem in the constant-free case.

While, by definition, GFU and \GFUTG{} disallow equality (and including unrestricted equality would lead to undecidability \cite{RS18}), we note that adding equality statements of the form $x=c$ to GFU and equalities guarded by transitive guards to \GFUTG{} can be done at no computational cost and would not affect our constructions at all.  The above additions nicely extend the expressive power of the logics. The first of them allows us, e.g., to express naturally the concept of nominals known
from description or hybrid logics, while with the second we can say that a transitive relation is actually an equivalence (cf.~\cite{DBLP:conf/lpar/KieronskiM20}, Section 5.1), which gives a chance to capture some scenarios from epistemic logics.

As a central open problem, it remains to clarify the decidability status of \GFTG{} and \GFUTG{} in the presence of constants.
We assume the resulting fragments will still be decidable. Obviously a lower complexity bound for \GFUTG{} with constants, inherited from \GFU, is \textsc{N2ExpTime} and hence harder than the constant-free case (under standard complexity-theoretic assumptions).

\section*{Acknowledgments}
E.K.~is supported by Polish National Science Centre grant No 2016/21/B/ST6/01444.
S.R.~is supported by the European Research Council through the ERC Consolidator Grant 771779 (DeciGUT).



%\bibliographystyle{IEEEtran}
\bibliographystyle{plain}
\bibliography{mybib,references}

\appendices
\section{Comments on the external procedures}

\noindent
{\bf Thm.~\ref{t:gfsize}.} 
     Inspecting the proof of Thm.~1.2 in \cite{BGO14}, we
		see that the size of a minimal finite model of a normal form formula over a signature $\sigma$ can be bounded by $|\str{J}|^{\mathcal{O}(w)}$ for some structure
		$\str{J}$ whose size is bounded by the number of guarded atomic types over $\sigma$,
		where $w$ is the width of $\sigma$. 
		
		The number of atomic $k$-types over $\sigma$ is $2^{\mathcal{O}(r(k+u)^w)}$, where $r$ is the number of relation symbols and $u$ the number of
		constants in $\sigma$. Of course, every guarded type is a $k$-type for some $k \le w$, so the number of guarded types
		is $w\cdot2^{\mathcal{O}(r(w+u)^w)}$, which is  $2^{\mathcal{O}(r(w+u)^w)}$. Thus the size of model is 
		$2^{\mathcal{O}(r(w+u)^{\mathcal{O}(w^2)})}$.
		As each of $r, w, u$ is bounded by $|\sigma|$ the claim follows.
	
\medskip\noindent
{\bf Thm.~\ref{t:gfcomp}.}  
The algorithm from \cite{Gra99} (page 1731) is an alternating algorithm.
The algorithm stores two guarded types and a counter counting up the the total number possible guarded types. That is it needs  $2^{\mathcal{O}({r(w+u)^w})}$ space. 
Using the classical simulation of alternating Turing machines by deterministic ones from \cite{CKS81} we get an algorithm
working in time $2^{\mathcal{O}(r(w+u)^w)} \cdot \mathcal{O}(n)$, where $n$ is the length of the input formula. 

\medskip\noindent
{\bf Thm.~\ref{t:gfttgsize}.} 
Let $\varphi$ be a normal form formula in \GFtTG{} over a signature $\sigma$. Denote $L$ the number of $1$-types over $\sigma$ ($L=2^{|\sigma|}$), $h$ the number of the $\forall \exists$-conjuncts of $\varphi$ and $k$ the number of transitive relations in $\sigma$.

In \cite{KT18} an important role is played by the parameter $M_\varphi$. It is defined on page 14  as $M_\varphi=3L|\varphi|^3$, but a closer inspection of the proof of Lemma 3.5 (iii), where the role of $M_\varphi$
is revealed, shows that it is sufficient
to take $M_\varphi=3Lh^3$.

Looking at page 27 of \cite{KT18} we see that the domain of the small finite model for $\varphi$ which is constructed there is of size bounded by
$$(2L+1) \cdot 4 \cdot h \cdot k \cdot T \cdot F$$
where $T$ is the number of the so-called \emph{enriched}-$M_\varphi$-\emph{counting types} and $F$ is a bound on the size of small models for the 'symmetric' part of $\varphi$.

$T$ is bounded by $2^L \times 2^L \times L^{M_\varphi+1}$, as each enriched-$M_\varphi$-counting type is determined by two subsets $\mathcal{A}$ and
$\mathcal{B}$ of $1$-types and a function which for each $1$-type returns a number from the range $\{0, \ldots, M_\varphi \}$ (see Def. 6.1). 
Substituting $3Lh$ for $M_\varphi$ we see that $T$ is bounded exponentially in $L$ (so, doubly exponentially in $|\sigma|$) and $h$. 

Concerning $F$, it is actually a bound on the size of the structure constructed in the proof of Lemma 5.7, that is  $3h (XL)^k$, 
where $X$ is the maximal value of a variable in some minimal solution for the system of linear inequalities constructed on page 20.
By Lemma 5.6  $X \le N \cdot (N M_\varphi)^{2N+1}$, for $N$ being the number of inequalities in the
system, which is bounded by $3L$. Summarizing, $F$ is not greater than $3h (3L^2 \cdot (9L^2 h^3)^{6L+1})^k$.

Gathering the above estimations we get that the size of the finite model constructed is exponential in the number of $1$-types (and thus 
doubly exponential in the size of the signature), in $h$, and in $k$. Taking into account that  $k \le |\sigma|$ the claim follows.

\medskip\noindent
{\bf Thm.~\ref{t:gfttgcomp}.}
Again, the algorithm from \cite{KT18} (page 30) is alternating. Assume that the notation is as in the previous paragraph.
What the algorithm stores is:
\begin{itemize}
\item a counter counting up to $kT$,
\item a collection of at most $kL$ enriched-$M_\varphi$-counting types
\end{itemize}
That is, the total space required is polynomial in the length of the input formula and exponential
in $L$ and $h$. Simulating alternating Turing machines by deterministic ones as previously we get an algorithm for normal form formulas working in time bounded polynomially in the size of the input and exponentially
in the number of $1$-types (doubly exponentially in the size of the signature) and the $\forall \exists$-conjuncts. 

%===========================================================================================
\section{Details on the FMP for GFU with constants} \label{a:constants}
Given a structure $\str{A}$ interpreting a signature $\sigma$ consisting of relation symbols and  constants we call the subset $\hat{A} \subseteq A$  consisting of all the elements interpreting the constants of $\sigma$ the \emph{named part} of $\str{A}$. The \emph{unnamed} part is defined as $\check{A}  := A \setminus \hat{A}$.
A $1$-type is \emph{named} if contains $x_1=c$ for some constant $c$ and \emph{unnamed} otherwise.
	
	We first redefine the notions of disjoint unions and doublings of structures.
 Let $(\str{A}_i)_{i \in \cI}$ be a family of $\sigma$-structures having disjoint unnamed parts and sharing the same named part; call such a family \emph{harmonized.} Their \emph{harmonized union} is the structure $\str{A}$ with
domain $A=\hat{A}_t \cup \bigcup_{i \in \cI} \check{A}_i$, where $t$ is chosen as any element of $\cI$, 
such that for all $i$ we have that $\str{A} \restr \hat{A}_t \cup \check{A}_i$ is isomorphic to $\str{A}_i$, and for any tuple $\bar{a}$ containing elements
from at least two different $\check{A}_i$ and any relation symbol $P \in \sigma$ of arity $|\bar{a}|$ we have $\str{A} \models \neg P(\bar{a})$. 

\begin{lemma}\label{l:disunion2}
Let $\varphi$ be a \GF{} normal form formula over a signature  consisting of relation symbols and constants. The harmonized union of any harmonized family of models
of $\varphi$ is also its model. 
\end{lemma}

Let $\str{A}_-$ be a $\sigma$-structure. Its \emph{harmonized doubling} is the structure $\str{A}$
with domain $A:=\hat{A}_- \times \{ 0 \} \cup \check{A}_- \times \{0, 1\}$ in which 
for each $P \in \sigma$ we have $\str{A} \models P[(a_1, \ell_1), \ldots, (a_k, \ell_k)]$ iff $\str{A}_- \models P[a_1, \ldots, a_k$] for all $a_i \in A_-$ and
  $\ell_i \in \{0, 1 \}$ if $a_i \in \check{A}_-$ and $\ell_i=0$ if $a_i \in \hat{A}_-$.
\begin{lemma}\label{l:doubling2}
Let $\varphi$ be a normal form \GF{} or \GFU{} formula which does not use equality outside guards and let $\str{A}_-$ be a model of $\varphi$. Then 
its harmonized doubling $\str{A}$ is still a model of $\varphi$.
\end{lemma}

Let us now fix a \GFU{} sentence $\varphi$ in normal form, without equality over a signature $\sigma$ consisting of relation symbols and
constants and let $\str{A}$ be a $\UU$-biquitous model of $\varphi$. Our goal is to build a finite $\UU$-biquitous model $\str{A}'$
of $\varphi$.

\subsection{Preparing building blocks}

We construct $\varphi^*$ precisely as in the case without constants.
We remark that this time each $\alpha \in \AAA$ fully specifies the substructure
consisting of the given element and the named part of the structure.

It is clear that $\varphi^*$, treated as a \GF-formula, is satisfiable. In fact, $\str{A}$ is its model.
Thus, by the finite model  property for \GF{}, 
it also has a finite (not necessarily $\UU$-biquitous) model. We take such a finite model  $\str{C}_- \models \varphi^*$, 
and let $\str{C}$ be its harmonized doubling.
As  $\varphi^*$ does not use equality,  by Lemma \ref{l:doubling2} we have that $\str{C} \models \varphi^*$.

Recalling the definition of indistinguishable elements we adapt Claim \ref{c:uconnected} to our current setting.
\begin{claim}\label{c:uconnected2}
For any pair of {\bf unnamed} $1$-types $\alpha, \alpha' \in \AAA$ there is a pair of their distinct realizations $a, a'$ in $\str{C}$
such that $\str{C} \models \UU[a, a'] \wedge \UU[a',a]$. Moreover, if $\alpha=\alpha'$, then we even find indistinguishable $a, a'$ with that property.    
\end{claim}

\medskip
As previously we build  yet another model $\str{B} \models \varphi^*$,  as the harmonized union of five  copies of $\str{C}$. 
Letting $K=|\check{C}|$, we assume that  the unnamed part of $\str{B}$ is $B:=\{1, \ldots, 5K \}$; and that for $m=0,\ldots, 4$ 
 the structure on $\hat{B} \cup \{mK+1, \ldots, mK+K\}$ is isomorphic to $\str{C}$. 

\subsection{$\boldsymbol{\UU}$-saturation} 

We now build a finite sequence of finite structures $\str{A}_0$, $\str{A}_1, \ldots, \str{A}_f$,
each of them being a model of $\varphi^*$ and the last of them being a desired $\UU$-biquitous model $\str{A}'$
of $\varphi^*$ (and thus also of $\varphi$).

The domains of all these structures will be identical.
\begin{align}
\nonumber
A_i= \hat{B} \cup ( \check{B} \times \{1, \ldots, 5K \} \times \{1, \ldots, 5K \} ). 
\end{align}

We will view each of the $\str{A}_i$ as a $5K \times 5K$ table containing unnamed parts plus
the shared named part.

The initial structure $\str{A}_0$ is defined 
as the harmonized union of $(5K)^2$ copies of $\str{B}$.
Namely,  for each  $k, \ell \in \{1, \ldots, 5K\}$ we make 
$\str{A}_0 \restr \hat{B} \cup (\check{B} \times \{k \} \times \{\ell\})$ isomorphic to $\str{B}$ 
(via the isomorphism working as the identity on $\hat{B}$ and as the natural projection $(b,k,\ell) \mapsto b$ on the unnamed elements). 
By Lemma \ref{l:disunion2} we have that $\str{A}_0 \models \varphi^*$.

\medskip\noindent
{\bf Some notation.}  We adapt our notation.  For each
$k,l$ we denote by $\str{B}_i^{k,\ell}$ the structure consisting of the common named part and the unnamed part in the cell $(k,\ell)$ of $\str{A}_i$, that is the structure $\str{A}_i \restr \hat{B} \cup (\check{B} \times \{k \} \times \{\ell \})$.
We recall that $\str{B}_0^{k,\ell}$ is isomorphic to $\str{B}$.
Further, for $m=0, \ldots, 4$, we denote by $\str{C}_i^{k,\ell,m}$ the structure $\str{B}_i^{k,\ell} \restr \hat{B} \cup \{mK+1, \ldots, mK+K \} \times \{k \} \times \{\ell \}$.
We recall that each $\str{C}_0^{k,\ell,m}$ is isomorphic to $\str{C}$.

\medskip\noindent
{\bf Entry elements and their use.}
This time only members of the unnamed parts are entry elements.
For any $1\le k,\ell \le 5K$, let $\alpha^k=\type{\str{B}}{k}$ and $\alpha^\ell=\type{\str{B}}{\ell}$. 
Note that $\alpha^k$ and $\alpha^{\ell}$ are unnamed.

For each such pair $k, \ell$ we now choose a pair of \emph{entry elements} for each of the five structures with unnamed parts in the cell $(k, \ell)$ of $\str{A}_0$, that is for the structures $\str{C}_0^{k,\ell,m}$ ($m=0, 1, \ldots, 4$).

By Claim \ref{c:uconnected2}, there are distinct elements $e_1, e_2 \in C$ such that $\str{C} \models \alpha^k[e_1]
\wedge \alpha^\ell[e_2] \wedge \UU[e_1, e_2] \wedge \UU[e_2,e_1]$ and if $\alpha^k=\alpha^\ell$ then $e_1$ and $e_2$ are indistinguishable in $\str{C}$.
We choose the entry elements  $e_1^{k,\ell,m}$, $e_2^{k,\ell,m}$ to $\str{C}_0^{k,\ell,m}$ to be the corresponding copies of $e_1$ and $e_2$
in each of $\str{C}_0^{k,\ell,m}$. 

By $^+\str{C}_i^{k,\ell,m}$ we denote the structure with domain $C_i^{k,\ell,m} \cup \{b_1,b_2\}$ for some fresh unnamed elements $b_1, b_2$  such that
$^+\str{C}_i^{k,\ell,m} \restr C_i^{k,\ell,m} = \str{C}_i^{k,\ell,m}$ and 
for each $P \in \sigma$ and each tuple $\bar{a}$ containing at least one of $b_1, b_2$ we have $^+\str{C}_i^{k,\ell,m} \models P[\bar{a}]$ iff $\str{C}_0^{k,\ell,m} \models
P[\fh(\bar{a})]$, where $\fh$ is the function defined as $\fh(b_1)=e^{k,\ell,m}_1$, $\fh(b_2)=e^{k,\ell,m}_2$ and $\fh(a)=a$ for $a \in C_i^{k,\ell,m}$ 
In particular $^+\str{C}_i^{k,\ell,m} \models \alpha^k[b_1] \wedge \alpha^\ell[b_2] \wedge \UU[b_1, b_2] \wedge \UU[b_2, b_1]$.

\medskip\noindent
{\bf From $\str{A}_i$ to $\str{A}_{i+1}$.}
Assume now that the structure $\str{A}_i$ has been defined, for some $i \ge 0$, $\str{A}_i \models \varphi^*$.
If $\str{A}_i$ is $\UU$-biquitous then we are done. 
Otherwise let $b_1, b_2$ be a pair of elements in $A_i$ such that 
$\str{A}_i \models \neg \UU[b_1, b_2]$. 
Note that  $b_1, b_2$ must be unnamed. Indeed, as in the case without constants, our process does not modify the types of the guarded tuples. In particular, 
	in each of the $\str{A}_i$ the $1$-types are retained from $\str{A}_0$, where they are copied from the model $\str{C}$ of $\varphi^*$. By the conjunct 
	(\ref{nf2}) of $\varphi^*$ all the $1$-types in $\str{A}_0$ belong to $\AAA$, the set of $1$-types realized in the $\UU$-biquitous structure $\str{A}$. 
	 Thus,
	if, say, $b_1$ would interpret a constant $c$ then, as the $1$-type of $b_2$  must contain $\UU(x,c) \wedge \UU(c,x)$ (since this $1$-type belongs to $\AAA$), we would have that $\str{A}_i \models \UU[b_1, b_2] 
	\wedge \UU[b_2, b_1]$.

For $s=1,2$ let $k_s, l_s, n_s$ be such that  $b_s$  is the $n_s$-th element of  the unnamed part of $\str{B}_i^{k_s,l_s}$.
Let us choose $t \in \{0, \ldots, 4\}$ such that $\str{C}_i^{n_1, n_2, t}$ does not
contain the $k_1$-th, $l_1$-th, $k_2$-th or $l_2$-th element of the unnamed part of $\str{B}_i^{n_1, n_2}$.
Such a $t$ must exist by the pigeon hole principle. We make the structure $\str{A}_{i+1} \restr
C_i^{n_1, n_2, t} \cup \{b_1, b_2\}$ isomorphic to $^+\str{C}_i^{n_1, n_2, t}$.
The rest of the structure $\str{A}_i$ remains untouched.

\medskip\noindent
{\bf Correctness.} The correctness of the construction can be proved in the same vain as in the case without constants.
In particular Claims \ref{c:1types}, \ref{c:noconflicts}, \ref{f:retaining} and \ref{c:correctness} remain true with
literally no changes and their proofs require only routine adjustments concerning the  division of
the domains into named/unnamed parts. That is the whole process eventually ends in $\UU$-biquitous model of $\varphi^*$.

\end{document}